\documentclass[useAMS,usenatbib]{mn2e}
\usepackage{txfonts}
\usepackage[latin1]{inputenc}
\usepackage[T1]{fontenc}
\usepackage{mathptmx}
\usepackage[scaled]{helvet}
\usepackage{courier}
\usepackage{aecompl} 
\usepackage{longtable}  
\usepackage{rotating}
\usepackage{natbib}
\usepackage{graphicx}
\usepackage{graphics}
\usepackage{psfrag}
\usepackage{amssymb}
\usepackage{multicol}
\usepackage{multirow}
\usepackage{lscape}
\usepackage{color}
\usepackage{soul}
\usepackage{array}
\usepackage{longtable}
\bibpunct{(}{)}{;}{a}{}{,}

\def\M{$\mathfrak{M}_{\odot}$}

\def\R21{$R_{21}$}
\def\r31{$R_{31}$}
\def\21f{$\phi_{21}$}
\def\31f{$\phi_{31}$}

\def\lt{LiTE}
\def\oc{\textit{O$-$C}}
\def\Blazhko{Bla\v{z}ko}

\author[K. T. Smith et al.]{
Keith T. Smith,$^{1}$\thanks{E-mail: mn@ras.org.uk (KTS)}
A. N. Other,$^{2}$
Third Author$^{2,3}$
and Fourth Author$^{3}$
\\
$^{1}$Royal Astronomical Society, Burlington House, Piccadilly, London W1J 0BQ, UK\\
$^{2}$Department, Institution, Street Address, City Postal Code, Country\\
$^{3}$Another Department, Different Institution, Street Address, City Postal Code, Country
}

\title[Cyclic variations in \oc~diagrams of RR Lyrae stars]{Cyclic variations in $\textrm{\oc}$~diagrams of field RR Lyrae stars\\ as a result of LiTE}

\author[J. Li\v{s}ka, M. Skarka, M. Zejda, Z. Mikul\'{a}\v{s}ek, \& S. N. de Villiers]{
J. Li\v{s}ka$^{1}$\thanks{jiriliska@post.cz}, M. Skarka$^{1,2}$, M. Zejda$^{1}$, Z. Mikul\'{a}\v{s}ek$^{1}$, \& S. N. de Villiers$^{3}$
\\
$^{1}$Department of Theoretical Physics and Astrophysics, Masaryk University, Kotl\'a\v{r}sk\'a 2, 611\,37 Brno, Czech Republic\\
$^{2}$Konkoly Observatory, Research Centre for Astronomy and Earth Sciences, Hungarian Academy of Sciences, H-1121 Budapest, Konkoly Thege Mikl\'{o}s
\\ 
\'{u}t 15-17, Hungary
\\
$^{3}$Private Observatory, 61 Dick Burton Road, Plumstead, 7800 Cape Town, South Africa
}

\begin{document}

\date{Received ..., 2016; }

\pagerange{\pageref{firstpage}--\pageref{lastpage}} \pubyear{2016}

\maketitle

\label{firstpage}

\begin{abstract}
This paper presents an extensive overview of known and proposed RR Lyrae stars in binaries. The aim is to revise and extend the list with new Galactic field systems. We utilized maxima timings for eleven RRab type stars with suspicious behaviour from the GEOS database, and determined maxima timings from data of sky surveys and our own observations. This significantly extended the number of suitable maxima timings. We modelled the proposed Light Time Effect (\lt) in \oc~diagrams to determine orbital parameters for these systems. In contrast to recent studies, our analysis focused on decades-long periods instead of periods in the order of years. Secondary components were found to be predominantly low-mass objects. However, for RZ Cet and AT Ser the mass of the suspected companion of more than one solar mass suggests that it is a massive white dwarf, a neutron star or even a black hole. We found that the semi-major axes of the proposed orbits are between 1 and 20 au. Because the studied stars belong to the closest RR Lyraes, maximal angular distances between components during orbit should at least be between 1 and 13 mas and this improves the chance to detect both stars using current telescopes. However, our interpretation of the \oc~diagrams as a consequence of the \lt~should be considered as preliminary without reliable spectroscopic measurements. On the other hand our models give a prediction of the period and radial velocity evolution which should be sufficient for plausible proof of binarity.
\end{abstract}
  
%________________________________________________________________

%*************************************************************************
\section{Introduction}\label{introductionsec}
%*************************************************************************

RR Lyrae stars play a crucial role in many astrophysical disciplines due to	their unique characteristics. Since they are evolved, luminous, population II stars with an easily recognizable light curve, they are used for the testing of evolutionary models \citep[e.g.][]{lee1990,pietrinferni2006} and research on the evolution, dynamics, and formation history of the Milky Way galaxy \citep{catelan2009,drake2013,stetson2014}. Because they are detectable over long distances, in combination with a known relationship between metallicity, period, and luminosity they represent suitable distance indicators also at extragalactic scales \citep[][]{catelan2008,fiorentino2010}. However, the location on the horizontal branch (HB) and the luminosity of a particular RR Lyrae star also depend on its mass. This parameter has been estimated only indirectly from evolution models and pulsation characteristics. Therefore independent determination of the mass of these stars is a key task that is sorely needed. If an RR~Lyrae would be in a binary system with known orbital parameters, its mass could be determined very accurately. Thus finding such systems is crucial.

Unfortunately, only very few RR Lyrae binary stars have been unambiguously identified so far \citep[e.g. the field star TU UMa in][]{wade1999,liska2015}. Moreover, the overwhelming majority of these stars are only candidates without definite confirmation of their binarity. This low number of binaries among RR Lyraes is surprising, because RR Lyrae variables are the most numerous class of variables \citep[VSX,][]{watson2006}, and recent compilations and catalogues with pulsating stars in binaries list many hundreds of systems with other types of pulsating stars \citep[e.g.][]{szatmary1990,szabados2003a,szabados2003b,zhou2010}. Well, what could be the reason?

Some possible reasons were discussed, for example, in the classical review of RR Lyrae stars provided by \citet{smith1995}. Since RR~Lyraes are evolved stars, their possible companion would be in the form of a degenerated remnant, another HB star, a red dwarf (in the case of main-sequence star), or a substellar object. Thus, in most cases, the companion would be significantly less luminous and smaller than the RR Lyrae component itself. They probably would be only single-line spectroscopic binaries dominated by the RR Lyrae component, and possible eclipses would be shallow and detectable with difficulty. As it was noted by \citet{soszynski2009} the mass transfer in close binaries would certainly affect an RR Lyrae progenitor. Therefore, if we want to discover a binary system containing a classical RR Lyrae, i.e. the one that has evolved without any interference from a companion, we should search for detached systems with wide orbits and with periods longer than a few hundreds of days, rather than for short-period binaries. All these facts would cause problematic detection due to tiny changes over years, or decades.

Recent research revealed that the situation is not as hopeless as previous studies suggested. Except for a strange object resembling RR Lyrae-like pulsations, identified by \citet{soszynski2011} in the Galactic bulge and studied by \citet{pietrzynski2012}, \citet{li2014}, on the basis of ultra-precise measurements by the {\it Kepler} space telescope, found tiny cyclic variations in the period (Light Time Effect -- \lt) of FN Lyr and V894~Cyg that are very likely caused by a brown dwarf companion. More recently, \citet{hajdu2015} identified 20 RR~Lyrae candidates in binary systems in the Galactic bulge using the same method. Additionally, \citet{hajdu2015} proposed that about 4\,\% of all RR~Lyraes reside in binary systems. The results of \citet{li2014} and \citet{hajdu2015} clearly demonstrated that studying cyclic period variations in these overluminous objects is much more efficient in revealing binaries than classical methods, i.e. detection of variations in position, variations in radial velocities (hereafter RV), or detection of eclipses. These results are very encouraging and hold the promise for more candidates to be revealed soon.

In our study we focussed on field RR Lyraes. We scanned the literature and the GEOS RR Lyr database\footnote{http://rr-lyr.irap.omp.eu/dbrr/dbrr-V1.0\_0.php} \citep[\textit{Groupe Europ\'{e}en d\textquoteright Observations Stellaires}, ][]{boninsegna2002,leborgne2007} to choose candidates with cyclic changes in their \oc~diagrams which can be interpreted as a manifestation of an unseen companion. Since there is no review of field stars suspected of binarity with reference to various methods currently available, we widely discuss particular targets in Sect.~\ref{rrbinarysec}. As we searched for manifestations of the \lt~in \oc, effects which can also affect the appearance of \oc~diagram are discussed (Sect.~\ref{rrocsec}). Except for \oc~values from the GEOS database, maxima timings from data of various sky surveys and from our own observations were determined (Sect.~\ref{datasec}). For modelling of the \lt~we used a new code introduced in \citet{liska2015}. Results for several targets are in Sect.~\ref{resultssec} and are generally discussed in Sect.~\ref{discussionsec}.

%*************************************************************************
\section{Overview of RR Lyrae stars suspected to be in binary systems}\label{rrbinarysec}
%*************************************************************************
Various types of binary systems with RR Lyrae stars, including known candidates, are discussed below. As a reference we prepared a comprehensive list \textit{RRLyrBinCan} of these objects, including false-positives, which is available at the CDS and in a regularly updated on-line version\footnote{http://rrlyrbincan.physics.muni.cz/}.
%_________________________________________________________________________
\subsection{Eclipsing systems}\label{rreclsubsec}

This type of binary stars is extremely important, because analysis of their light and RV curves allows for the precise determination of absolute values of their basic physical parameters. A few RR Lyrae stars were proposed to be bound in such systems. Nevertheless, the eclipses are in most cases based on unique observations without any later confirmation. This could either be some instrumental bias or it could be a consequence of a long-period binary with very rare eclipses.

\citet{fitch1966} reported anomalous behaviour of VX~Her in one night's observation data when the star was about 0.7\,mag fainter than usual in a phase of minimum-light. Unfortunately, no repetition of this event has since been reported.

Similar behaviour was noticed in RW Ari, an RRc type star, by \citet{wisniewski1971}. He observed this star during 19 nights in 1966. On two nights RW Ari was about 0.6\,mag and on another night about 0.1\,mag fainter than usual in the \textit{V}-band. He proposed these dips in brightness to be primary and secondary minima of an eclipsing binary. After subtraction of the average light curve he obtained a typical light curve of an eclipsing binary -- a very convincing proof of his hypothesis (his Fig.~2). According to changes in $B\!-\!V$ and $U\!-\!B$ indices he suggested that the companion is a blue giant or B-type star, which is more luminous than RW Ari itself. \citet{wisniewski1971} determined the period of eclipses to be a bit longer than 3\,d. Another decrease in brightness (0.2\,mag, supposed secondary minimum), which confirmed the period of eclipses, was reported by \citet{woodward1972}, who analysed data from \citet{detre1937}. Spectroscopic observations made by \citet{abt1972} showed a very strong difference in RV measured at the same pulsation phase, which was interpreted as a consequence of binarity. 

However, \citet{edwards1978} measured RW Ari on 35 nights during the seasons of 1976 and 1977 and disproved all formerly noted statements about this star. In his observations no indication of eclipses was detectable, and the colour of RW Ari was found to be very similar to that of other RRc-type stars. He only suggested that RW Ari could possibly show \Blazhko~modulation with periods of either 38 or 114\,d. In addition, all following attempts to detect eclipses failed. 

It is worth mentioning that reports on eclipses of VX Her and RW Ari were based on observations performed by one group of observers, at the same time, and using the same equipment and methods whereby no comparison star was used \citep[only standard stars, ][]{fitch1966,bookmeyer1977}. Therefore, the possibility of a false detection could not be ruled out in these two cases \citep[more in][]{liska2016}.

V80 UMi, located in the Ursa Minor Dwarf galaxy, is the other candidate for an RR Lyrae pulsating star in an eclipsing system. The binary nature, proposed by \citet{kholopov1971}, is based on photographic measurements presented in \citet{vanagt1968}. Kholopov found a possible eclipsing period close to 2\,d and a pulsation period of about 0.5\,d, but both values are dubious, because they are close to the integer ratio of one day. In addition the brightness depressions were detected only on 10 plates of three consecutive nights. Before and after that nothing special was observed. V80 UMi is very faint with variations in $B$-band of 19.20\,--\,20.25\,mag, and this was probably the reason why only one additional dataset was obtained for this star \citep{nemec1988}. Their photographic measurements do not support eclipses. Nevertheless, they improved the pulsation period to 0.498746\,d and proposed the type of the variable to be an anomalous cepheid.

The systematic search for an RR Lyrae eclipsing system in the Galactic bulge (3256 fundamental-mode RR Lyraes from MACHO project) was unsuccessful \citep{richmond2011}. Therefore, the only currently known RR Lyrae-like star in an eclipsing binary, which was confirmed spectroscopically, is OGLE-BLG-RRLYR-02792 located in the Galactic bulge \citep{pietrzynski2012}. Unfortunately, this low-mass star (0.26\,$\mathfrak{M}_{\odot}$) with a pulsation period of 0.6275\,d very likely only behaves similarly to an RR Lyrae pulsator because of its evolution in a close binary system\footnote{The secondary, more massive component, is at the beginning of its red-giant evolutionary phase.} -- the current orbital period is 15.24\,d and the semi-major axis is 32\,R$_{\odot}$. The reasons for excluding this star from the classical RR Lyrae group were discussed by \cite{pietrzynski2012} and \citet{smolec2013}. They found that the bump in the ascending branch of the RV curve, Fourier parameter $\phi_{31}$ calculated with theoretical models, and a very high rate of period-decrease rules OGLE-BLG-RRLYR-02792 out as a regular RR Lyrae star. Therefore \citet{pietrzynski2012} defined it as a prototype of a new class of variables, called Binary Evolution Pulsators. Another object, OGLE-LMC-RRLYR-03541 in the LMC \citep{soszynski2009}, probably also belongs to this class \citep{hajdu2015b}. However, neither has the possibility that the star is a blend of two objects (eclipsing binary and RR Lyrae star) been rejected yet.

%_________________________________________________________________________
\subsection{Systems with \lt}\label{rrlitesec}

Binarity of a system with a pulsating star can be unveiled even without eclipses through periodic changes in the times of maximum light occurring in the pulsating component, which is caused by the orbit of an unseen companion. This is usually examined by using \oc~diagrams (more in Sect.~\ref{rroclitesubsec}).

A few RR Lyrae stars were proposed to be in binary systems due to \lt. Probably the most well-known example is TU~UMa. This star shows sinusoidal propagation in its times of maximum light with a period of about 23 years, which is superimposed on a parabolic trend caused by stellar evolution. The idea of binarity of TU UMa was first noted by \citet{szeidl1986} and further analysed by \citet{saha1990}\footnote{\citet{saha1990} and \citet{wade1999} analysed also systematic shifts in RVs.}, \citet{kiss1995} and \citet{wade1999}.

Original measurements, current observations available in large sky surveys, and observations from photographic plates allowed us to perform very precise reanalysis of the \oc~of TU UMa \citep{liska2015}. It was found that TU~UMa very likely comprises a pair with a low-mass (minimum value 0.33\,$\mathfrak{M}_{\odot}$), low-luminous star with orbital period of 23.4\, years. The binary assumption was supported by the analysis of available RV measurements.

XZ~Dra shows in its \oc~diagram a similar period of cyclic variations of about 7200\,d (19.7\,yr) together with a slowly increasing pulsation period as analysed by \citet{jurcsik2002}. They estimated orbital parameters of the possible binary system which allowed them to describe the \oc~diagram and variation in RV measurements. They argued that another mechanism, e.g. a magnetic cycle, can play a role.

Another very interesting object showing \lt-like behaviour is BE~Dor\footnote{BE Dor = MACHO* J050918.712-695015.31.}, studied by \citet{derekas2004}. If the binary nature of this RRc star is correct, the RR Lyrae component would orbit with a period of 8 years around the center of mass together with a black hole with a mass of about 60\,$\mathfrak{M}_{\odot}$. Although they argue against this interpretation, and prefer some sort of magnetic cycle instead of the binary explanation, the situation with this peculiar object is still unclear.

In the last year, \citet{li2014} analysed tiny changes in the times of maximum of FN Lyr and V894 Cyg based on data from the {\it Kepler} telescope. They found that these stars are both accompanied by a low-mass object, probably a brown dwarf or giant planet, with orbital periods of 795\,d (FN Lyr) and 1084\,d (V894 Cyg). It is very likely that more such binaries will be unveiled in precise space measurements \citep[e.g.][]{guggenberger2014}.

Very recently, twenty binary candidates were identified among OGLE bulge fundamental-mode RR Lyraes by \citet{hajdu2015}. They started analysis from 1952 of well-observed stars with baselines of up to 17 years. From these results, they estimate that $\gtrsim 4$\,\% of all RR Lyrae stars belong to binary systems. 

Analysis of the long-term period behaviour of AV Peg, RS Boo, RR Leo \citep{olah1978,szeidl1986} and RZ Cet \citep{leborgne2007} also provided some indication of the binarity of these objects. In addition, \citet{leborgne2007} gave several other somewhat tentative examples with some peculiarity in their \oc~diagrams, which can possibly be a consequence of binarity (e.g. SX Aqr, X Ari, etc). Long-term monitoring, lasting a few decades, facilitated the identification of several RR Lyrae candidates in binary systems in globular clusters, e.g. M3, M5 or M53 \citep{szatmary1990,szeidl2011,jurcsik2012}.

%_________________________________________________________________________
\subsection{Spectroscopic binaries}\label{rrspecbinsec}

Spectroscopic binaries, discovered through variations in RV, comprise a large fraction of known binary systems among ordinary stars. Nevertheless, this statement, at least partially, does not apply to RR Lyrae stars, in which the situation is complicated due to pulsations. In addition, in detached systems, the pulsation component probably would dominate in the spectrum (reasons were mentioned in Sect.~\ref{introductionsec}). Thus, only long-term, high-precision spectroscopic observations, and application of spectral disentangling, could bring long-desired success.

So far the only known spectroscopic binary with an RR~Lyrae component is OGLE-BLG-RRLYR-02792 \citep{pietrzynski2012}, which is also an eclipsing binary (see Sect.~\ref{rreclsubsec}). This star is a double-line spectroscopic binary with the main component only mimicking an RR Lyrae type star. Concerning galactic field stars, TU~UMa is one of the most promising targets, in which possible RV changes should be observable during one decade \citep{liska2015}. Variation in RV was also analysed in XZ Dra \citep[Sect.~\ref{rrlitesec}, ][]{jurcsik2002}.

Several candidates for spectroscopic binaries were identified based on differences in systemic RV from different studies which achieve up to several tens of km\,s$^{-1}$. The list of suspected systems from \citet{fernley1997} contains CI~And, DM~Cyg, BK~Dra, XX~Hya, ST~Leo, CN~Lyr, and TU~UMa, and is complemented by TY~Aps, BX~Dra, and BX~Leo from \citet{solano1997}. DM Cyg and TY Aps show \Blazhko~modulation, which could produce a shift in RVs, and BX~Dra was re-classified as an eclipsing binary \citep{agerer1995}.

%_________________________________________________________________________
\subsection{Visual binaries}\label{rrvisbinsec}

Although the great distances to RR Lyrae stars make visual pairs hardly detectable, this type of binaries should be mentioned at least for completeness. The upper limit for the separation of gravitationally bound objects is probably not known. Nevertheless, from a catalogue of 3139 orbits from \citet{malkov2012}, we found that 99.3\,\% stars in 2869 binaries have semi-major axes shorter than 1000\,au. For example the prototype RR~Lyrae itself \citep[the closest RR Lyrae star, parallax $\pi = 3.46$\,mas,][]{vanleeuwen2007} as a hypothetical system with semi-major axis of 1000\,au would have a component with a maximum angular distance of $3.46''$. This illustrates the difficulty in finding visual systems among RR Lyraes, because possible visual systems would be much closer and changes in position angle undetectable on reasonable time scales. The chance for discovery of such a system without adaptive optics or interferometry is negligible.

%_________________________________________________________________________
\subsection{Other types of binarity}\label{rrothersec}
In the second half of the twentieth century, AR Her and BB Vir were found to be significantly hotter at minimum light than other RRab variables \citep{preston1959,sturch1966,fitch1966,bookmeyer1977}. Later, this discrepancy was discussed and explained by \citet{kinman1992} and \citet{fernley1993} as a possible consequence of binarity. They considered the second component to be a blue, hot star.

Both BB Vir and AR Her are known as \Blazhko~stars, but the modulation nature of BB Vir is somewhat questionable \citep{skarka2014}. \citet{kinman1992} analysed photometric observations of BB Vir in detail using $(B\!-\!V)$ colour and also light curve characteristics. They found that BB Vir has a lower $(B\!-\!V)$ than other RRab stars, and that the amplitude of its light curve is smaller than it should be for the given colour. After their correction of the BB Vir light (assumed to be similar as for other RRab stars), they suggested the unresolved component to be a hot HB star with similar physical characteristics as BB Vir itself, which lies on the blue edge of the instability strip. Based on ultraviolet spectra, \citet{fernley1993} more closely specified the temperature of the supposed companion to 7900\,K. He also suggested an alternative explanation of BB Vir to be an RRc star with an unusually long period.

\citet{borkowski1980} explained the modulation of AR Her to be the consequence of non-linear superposition of the fundamental ($P_{1}=0.4700$\,d) and the second or third overtone ($P_{2}=0.233$\,d). \citet{kinman1992} speculated that, if the binary nature of AR Her is correct, the contribution of the second overtone to its modulation could actually be explained by the presence of an unresolved RRc star with a 0.233-day period rather than the consequence of beating between radial overtones. \citet{kinman1992} and \citet{fernley1993} nicely showed how expected photometric and spectroscopic characteristics of RR Lyrae stars would be affected by a possible companion. Despite all the discussed theories, a convincing proof of the binarity of BB Vir and AR Her is still missing.

%*************************************************************************
\section{Period changes of RR Lyrae stars in $\textrm{\oc}$~diagram}\label{rrocsec}
%*************************************************************************

Since the beginning of modern variable star research, the very simple and illustrative \oc~diagram has been used for the description of the period evolution of stars. Time of a given phase (usually maximum or minimum of the light changes), which is observed ($O$), is compared with predicted time calculated ($C$) on the basis of an accepted ephemeris. Subsequently the time difference \oc~is plotted against time or a cycle number $N$.

In an \oc~plot it is very easy to investigate long-term systematic trends, which can be e.g. monotonically linear (constant period) or parabolic (period change with a constant rate), periodic, as well as sudden, more or less irregular \citep[a review of \oc~diagram shapes can be found e.g. in][]{liska2015b}. In the case of RR Lyrae stars, the behaviour of their periods can be very complex with many possible reasons combined, sometimes with an unclear interpretation. Therefore, an RR Lyrae star showing \lt, without another independent confirmation, should only be considered to be a ``candidate'' binary system. Very illustrative examples of many \oc~diagrams can be found e.g. in \citet{jurcsik2001,jurcsik2012,szeidl2011,leborgne2007}.

%_________________________________________________________________________
\subsection{Evolutionary effects}\label{rrocevolsubsec}

Already at the beginning of the last century, \citet{eddington1918} pointed out that pulsating stars should undergo evolutionary period changes due to changes in their density. These changes are very slow, therefore observations spread out over many decades are needed. Fortunately, RR Lyraes have been observed for more than a century in some particular cases.

In RR Lyrae research, it is usual to express the period-change rates in $\beta=\dot{P}_{\rm puls}={\rm d}P_{\rm puls}/{\rm d}t$ in [d\,Myr$^{-1}$] or [ms\,d$^{-1}$], and $\alpha=\beta/P_{\rm puls}=(1/P_{\rm puls})\,\dot{P}_{\rm puls}$ [Myr$^{-1}$] parameters, for definition of $\dot{P}_{\rm puls} = 2\,a_{3}/P_{\rm puls}$, where $a_{3}$ is a third coefficient in the parabolic ephemeris of maxima timings \citep[e.g.,][]{leborgne2007}\footnote{Several authors \citep[e.g.][]{hajdu2015} used different definition of $\beta$-parameter -- their $\beta_{\rm their} = a_{3}$ and then $\beta_{\rm their} = 1/2\,P_{\rm puls}\,\beta$.}. Depending on their masses, chemical composition and evolutionary stage, RR Lyrae variables can exhibit both a period decrease and a period increase during their evolution on the HB \citep[see e.g. the theoretical works of][]{sweigart1987,lee1990}. When the evolution is blueward, the period of a star slowly decreases, while the evolution from blue to red implies a period-lengthening. Since the redward evolution is more rapid than towards the blue, the rate of change of the period is higher for period-lengthening than for period shortening. 

According to \citet{lee1990}, the majority of RR Lyrae stars are in the early stage of their HB evolution. Analysing a large sample of stars such as in globular clusters, \citet{lee1990} found that the average value of $\beta$ should be around zero or very slightly positive. This assumption was confirmed by \citet{lee1991}, who compared synthetic with the observed HB and found that $\beta$ depends on the type of HB. Another confirmation came with studies \citet{jurcsik2001,jurcsik2012} and \citet{szeidl2011}, that also investigated period changes of RR Lyraes in globular clusters.

Nevertheless, in all these studies, as well as e.g. in \citet{leborgne2007}, who analysed the \oc~of field stars, some peculiar cases with a very high positive or negative $\beta$ were reported. Stars with a large $\beta>0$ are considered to undergo the final episodes of their HB life, while RR Lyraes with the most rapid period-decrease are supposed to be in their pre-HB stadium \citep{sweigart1979,silva2008}. However, in several cases with the largest $\beta$, such high period changes cannot be explained by evolution theory. We suppose that for some of these stars the discrepancy between observation and theory could be caused by misinterpretation of their \oc~diagrams. In the case of a potential double star with a long orbital period, the corresponding \oc~could mimic a parabola when only a short time span of observation is available.

In a few objects the shape of the \oc~diagram is more complicated than only a simple parabola. A cubic ephemeris with significant third-order
terms was also sometimes calculated for these objects \citep[e.g., RS Boo, SU Dra, RW Cnc,][]{leborgne2007}. This approach is meaningless as was shown by \citet{mikulasek2013}. Evolutionary changes described using the power law, in which ${\rm d}P_{\rm puls}/{\rm d}t \sim P_{\rm puls}^{2-q}$, where $q$ is the deceleration parameter, the third-order term is negligible even in objects with extreme period shortening/lengthening. This implies more complex behaviour than simple evolutionary effects are present in these objects (e.g. binarity).

%_________________________________________________________________________
\subsection{\Blazhko~effect}\label{rrocblazhkosubsec}

Almost a half of all RRab stars \citep{jurcsik2009,benko2010} shows more or less periodic changes of the shape of their light curves, which is known as the \Blazhko~effect \citep{blazhko1907}. These changes manifest as changes in amplitude and in time of maximum light. Therefore, these changes can be reproduced and analysed using an \oc~diagram. An \oc~diagram of a \Blazhko~star shows a periodic pattern with an amplitude of maximally about 0.2\,d \citep[the currently known record holder is V445~Lyr with $\Delta$(\oc)$\,\sim 0.15$\,d,][]{guggenberger2012}. Periods of the \Blazhko~effect differ by several orders -- from 5 days to more than 5 years \citep{szczygiel2007}. In fact, the upper limit for the length of the \Blazhko~cycle is not known, and the greatest values detected for the period are governed by the length of the time-base. Thus the \Blazhko~effect can be easily misclassified with \lt.

\subsection{Erratic and other changes}\label{rrocerraticsubsec}
Except for periodic changes caused by the \Blazhko~effect, RR Lyraes undergo very sudden, random changes, which affect mainly \Blazhko~stars \citep[see e.g.][]{jurcsik2012,leborgne2007}. These abrupt, significant changes observed, for example, in XZ~Cyg \citep{bezdenezhnyi1988} or RR~Gem \citep{sodor2007}, are usually associated with instabilities during the final phase of the helium-burning phase or as a consequence of the mixing events at the convective-core edge \citep{sweigart1979}. Other possible explanations deal with changes in the gradient of the helium composition in the regions below the hydrogen and helium convection zones \citep{cox1988}.

Additional, not necessarily erratic changes, can be connected with hydromagnetic effects. They are probably the only way to explain the problem with high-amplitude cyclic variations detected in non-\Blazhko~stars which appears like peculiar \lt~caused by a very massive component \citep[a possible black hole with several tens of $\mathfrak{M}_{\odot}$, e.g. BE~Dor,][]{derekas2004}.

%_________________________________________________________________________
\subsection{Light Time Effect}\label{rroclitesubsec}

Probably the last possible way to affect the \oc~values of RR Lyrae stars, is through the effect of orbital motion in a binary system with appropriate parameters (large semi-major axis and inclination). The orbital motion of an RR Lyrae component will produce detectable changes of its times of maximum light. This is a consequence of the finite speed of light, and it is called the Light Time Effect (\lt), as already mentioned. When a pulsating star is in the closest part of its orbit to the observer, the time needed to reach the observer is logically shorter than when it is in other parts of the orbit. As the star revolves around the center of mass, the corresponding \oc~plot will change periodically. The shape of these changes depends on the orientation and eccentricity of the orbit. The amplitude of such an \oc~diagram directly provides information about the projected size of the orbit, and the period of these changes correlates with the orbital period. These parameters allow us to estimate the total mass of the system. A few systems suspected of \lt~were briefly discussed in Sect.~\ref{rrlitesec}.

\begin{center}
\begin{table}
\caption{Number of new times of maxima determined from individual projects (HIP -- Hipparcos), from our observations, and GEOS database.}
\centering
%\tiny
\scriptsize
%\footnotesize
\def\arraystretch{1.5}
\tabcolsep=1.3pt
\begin{tabular}{lccccccccccc}
\hline
Star	& ASAS-3	& CRTS	& DASCH	& HIP	& NSVS	& OMC	& SWASP	& Own & GEOS	& Interval\\
\hline
RS Boo	& 0			& 0		& 49		& 3				& 4		& 0		& 32		& 0		& 448	& 1900.0\,--\,2013.5 \\
RU CVn	& 0			& 5		& 24		& 0				& 6		& 0		& 43		& 0		& 114	& 1902.0\,--\,2013.3 \\
RZ Cet	& 15		& 8		& 0			& 3				& 9		& 4		& 0			& 0		& 100	& 1930.0\,--\,2013.8 \\
S Com	& 10		& 10	& 34		& 3				& 11	& 0		& 0			& 0		& 134	& 1901.2\,--\,2013.4 \\
SU Dra	& 0			& 0		& 0			& 5				& 5		& 0		& 0			& 0		& 235	& 1904.2\,--\,2013.5 \\
VX Her	& 13		& 4		& 0			& 4				& 9		& 0		& 0			& 6		& 243	& 1916.2\,--\,2014.6 \\
SS Leo	& 13		& 2		& 0			& 2				& 3		& 0		& 0			& 3		& 105	& 1901.3\,--\,2014.4 \\
AV Peg	& 9			& 0		& 0			& 3				& 4		& 0		& 0			& 0		& 502	& 1931.7\,--\,2013.6 \\
AT Ser	& 14		& 7		& 0			& 3				& 6		& 1		& 0			& 3		& 50 	& 1900.9\,--\,2014.5 \\
RV UMa	& 0			& 0		& 53		& 5				& 14	& 12	& 0			& 0		& 252	& 1897.3\,--\,2013.3 \\
BB Vir	& 5			& 7		& 38		& 2				& 5		& 6		& 0			& 0		& 46 	& 1902.1\,--\,2013.4 \\
\hline
\end{tabular}\label{Tab:NumberMaximatable}	
\end{table}
\end{center}

%*************************************************************************
\section{Data sources and target selection}\label{datasec}
%*************************************************************************

Since the GEOS RR Lyrae database \citep[][]{leborgne2007} is the most extensive archive containing times of maxima of RR Lyrae stars, it was used as the main source of data for this work. Targets from literature suspected of \lt~were supplemented by objects selected only on the basis of visual inspection of their \oc~diagrams from the GEOS web page. This means that only stars with suspect cyclic variations were selected (stars showing one or more cycles or almost one cycle dissimilar to a parabola). In addition, only stars with more than 10 maxima timings passed for further analysis from the entire GEOS database. After a closer inspection of the remaining \oc~diagrams only eleven RRab stars, which are discussed in Sect.~\ref{resultssec}, were fully analysed. From the preceding discussion it is evident that we focused on high amplitude and long-term variations in \oc~diagrams with good time-coverage. Therefore, our study is strongly affected by selection effects.

We did not use data marked as \textquoteleft pr. com.\textquoteright~from the GEOS database (the numbers of used maxima from the GEOS database are in the Table~\ref{Tab:NumberMaximatable}). To maximally extend the GEOS sample we re-analysed\footnote{From the reasons discussed in \citet{liska2015}.} data from sky surveys -- Hipparcos \citep{esa1997}, NSVS \citep{wozniak2004}, ASAS-3 \citep{pojmanski1997,pojmanski2002}, and gained new maxima from CRTS\footnote{http://crts.caltech.edu/}, DASCH \citep{grindlay2009}, SuperWASP \citep{pollacco2006,butters2010}, and OMC\footnote{http://sdc.cab.inta-csic.es/omc/index.jsp} \citep[see e.g][]{alfonso2010,zejda2011}. New maxima for the studied systems were determined from sparse data using the template fitting method described in \citet{liska2015}. The dataset with the best quality (good accuracy, phase coverage) was selected for each star for the template fit. Maxima from SuperWASP measurements (good time-resolution) were calculated using polynomial fitting. The determined times, including uncertainties, obtained directly from the least-squares method are available through CDS. The number of used maxima from particular surveys can be found in Table~\ref{Tab:NumberMaximatable}.

\begin{table*}
%\centering	
\begin{minipage}{175mm}
\begin{center}  
\caption{Identification and basic light characteristics of studied stars \citep[taken from The International Variable Star Index -- VSX,][]{watson2006}.}
%{\scriptsize
\def\arraystretch{1.5}
\tabcolsep=12.5pt
\begin{tabular}{llllllll}
\hline
Star & RA [$^{\rm h}$:$^{\rm m}$:$^{\rm s}$] & DEC [$^{\circ}$:$'$:$''$] & Other designation & $M_{0}$ [HJD] & $P_{\rm puls}$ [d] & Mag range $(V)$ \\ \hline
RS Boo & 14:33:33.21 & $+$31:45:16.6 & AN 14.1907 = HIP 71186 & 2441770.49 & 0.37733896 & 9.69\,--\,10.84\\
RU CVn & 13:59:33.28 & $+$31:39:04.3 & AN 6.1914\hspace{1.7cm}  & 2434483.467 & 0.5732449 & 11.36\,--\,12.48\\
RZ Cet & 02:28:32.44 & $-$08:21:30.1 & AN 36.1929 = HIP 11517 & 2433906.8920 & 0.5105981 & 11.245\,--\,12.240\\
S Com  & 12:32:45.63 & $+$27:01:45.4 & AN 83.1910 = HIP 61225 & 2440654.6410 & 0.58658722 & 10.88\,--\,12.12\\
SU Dra & 11:37:56.61 & $+$67:19:47.0 & AN 43.1907 = HIP 56734 & 2443902.0467 & 0.66042001 & 9.18\,--\,10.27\\
VX Her & 16:30:40.73 & $+$18:22:00.2 & AN 30.1917 = HIP 80853 & 2453470.2068 & 0.4553595 & 9.913\,--\,11.177\\
SS Leo & 11:33:54.49 & $-$00:02:00.0 & AN 69.1919 = HIP 56409 & 2441781.4090 & 0.6263351 & 10.420\,--\,11.584\\
AV Peg & 21:52:02.79 & $+$22:34:29.4 & AN 90.1931 = HIP 107935 & 2443790.3160 & 0.3903814 & 9.93\,--\,10.99\\
AT Ser & 15:55:40.44 & $+$07:59:18.7 & AN 46.1935 = HIP 77997 & 2453467.0077 & 0.7465617 & 11.013\,--\,11.890\\
RV UMa & 13:33:18.09 & $+$53:59:14.6 & AN 139.1907 = HIP 66122 & 2445075.511 & 0.46806 & 9.81\,--\,11.3\\
BB Vir & 13:51:40.78 & $+$06:25:51.4 & AN 23.1935 = HIP 67653 & 2439613.7760 & 0.4711069 & 10.732\,--\,11.450\\
\hline
\end{tabular}\label{Tab:BasicCharTable}
\end{center}
\end{minipage}  
\end{table*}

\begin{table*}
\centering
\begin{minipage}{175mm}
\caption{Pulsation and orbital parameters for studied systems with probable \lt$^{*}$.}
{\scriptsize
\def\arraystretch{1.5}
\tabcolsep=1.3pt
\begin{tabular}{lcccccccccccccccccc}
\hline
		 Star		& Model	& $P_{\rm puls}$			& $M_{0}$			& $10^{-11}\,\dot{P}_{\rm puls}$	& $\beta$		& $P_{\rm orbit}$		& $T_{0}$			& $e$			& $\omega$		& $A$			& $a_{1}\sin i$ 	& $f(\mathfrak{M})$		& $\mathfrak{M}_{\rm 2, min}$ 	& $K_{1}$ 		& $\chi_{\rm R}^2$	& $N_{\rm max}$\\
			&	& [d]						& [HJD]				& [d\,d$^{-1}$]				& [d\,Myr$^{-1}$]	& [yr]				& [HJD]				&			& $[^{\circ}]$		& [light day]		& [au]			& [$\mathfrak{M}_{\odot}$]	& [$\mathfrak{M}_{\odot}$]	& [km\,s$^{-1}$]	&			& \\
\hline
		 RS Boo$^{B}$	& 1	& 0.37733735$^{+12}_{-17}$		& 2448500.3193$^{+24}_{-17}$	& --					& --			& 147$^{+23}_{-15}$		& 2425500$^{+4500}_{-3300}$	& 0.150$^{+86}_{-25}$	& 220$^{+37}_{-30}$	& 0.0468$^{+78}_{-45}$	& 8.1$^{+1.3}_{-0.8}$	& 0.0246$^{+48}_{-28}$		& 0.264$^{+19}_{-15}$		& 1.660$^{+65}_{-35}$	& 1.082(61)		& 536\\
\vspace*{0.15cm} RS Boo$^{B}$	& 2	& 0.377337689$^{+26}_{-13}$		& 2435747.0346$^{+6}_{-25}$	& $+12.81\,^{+76}_{-19}$		& $+0.0468^{+28}_{-7}$	& 76.7$^{+0.6}_{-2.7}$		& 2455350$^{+350}_{-560}$	& 0.55$^{+22}_{-19}$	& 213$^{+11}_{-6}$	& 0.0129$^{+14}_{-3}$	& 2.23$^{+25}_{-4}$	& 0.00188$^{+71}_{-10}$		& 0.097$^{+12}_{-3}$		& 1.03$^{+26}_{-8}$	& 1.082(62)		& 536\\
\vspace*{0.15cm} RU CVn		& 1	& 0.573250314$^{+49}_{-89}$		& 2455343.4612$^{+64}_{-21}$	& --					& --			& 101.6$^{+3.1}_{-4.2}$		& 2448500$^{+3700}_{-3500}$	& 0.22$^{+15}_{-2}$	& 333$^{+39}_{-40}$	& 0.0495$^{+30}_{-65}$	& 8.6$^{+0.5}_{-1.1}$	& 0.061$^{+10}_{-19}$		& 0.392$^{+34}_{-55}$		& 2.58$^{+14}_{-23}$	& 1.03(10)		& 192\\
\vspace*{0.15cm} RZ Cet$^{B?}$	& 2	& 0.51061244$^{+28}_{-44}$		& 2441276.6248$^{+98}_{-60}$	& $-180.8\,^{+7.6}_{-6.9}$		& $-0.660^{+28}_{-25}$	& 75.4$^{+2.2}_{-4.0}$		& 2431530$^{+600}_{-610}$	& 0.35$^{+16}_{-7}$	& 105.2$^{+9.5}_{-8.8}$	& 0.082$^{+10}_{-11}$	& 14.1$^{+1.7}_{-1.9}$	& 0.49$^{+15}_{-15}$		& 1.15$^{+17}_{-25}$		& 5.94$^{+74}_{-30}$	& 1.13(12)		& 139 \\
\vspace*{0.15cm} S Com		& 1	& 0.586589092$^{+56}_{-33}$		& 2453796.55538$^{+93}_{-59}$	& --					& --			& 105.8$^{+6.2}_{-3.7}$		& 2442200$^{+2500}_{-2700}$	& 0.12$^{+13}_{-6}$	& 135$^{+22}_{-25}$	& 0.0329$^{+13}_{-6}$	& 5.70$^{+23}_{-10}$	& 0.0166$^{+14}_{-10}$		& 0.2241$^{+78}_{-55}$		& 1.617$^{+55}_{-34}$	& 1.05(10)		& 202\\
\vspace*{0.15cm} S Com		& 2	& 0.586589124$^{+87}_{-9}$		& 2435949.0067$^{+34}_{-20}$	& $-7.0\,^{+1.2}_{-4.1}$		& $-0.026^{+4}_{-15}$	& 90.1$^{+2.5}_{-7.8}$		& 2442700$^{+4500}_{-700}$	& 0.193$^{+96}_{-44}$	& 134$^{+62}_{-17}$	& 0.0228$^{+19}_{-34}$	& 3.96$^{+34}_{-58}$	& 0.0076$^{+17}_{-21}$		& 0.165$^{+13}_{-22}$		& 1.33$^{+12}_{-9}$	& 1.03(10)		& 202\\
\vspace*{0.15cm} SU Dra		& 2	& 0.66042124$^{+9}_{-12}$		& 2443902.0549$^{+23}_{-13}$	& $+16.6^{+0.7}_{-1.2}$			& $+0.0606^{+27}_{-44}$	& 70.7$^{+3.1}_{-3.6}$		& 2432100$^{+3700}_{-3300}$	& 0.24$^{+32}_{-1}$	& 109$^{+50}_{-47}$	& 0.0113$^{+21}_{-12}$	& 1.96$^{+36}_{-21}$	& 0.00150$^{+80}_{-47}$		& 0.089$^{+18}_{-13}$		& 0.85$^{+24}_{-9}$	& 1.118(91)		& 245\\
\vspace*{0.15cm} VX Her$^{B?}$	& 2	& 0.455366782$^{+31}_{-15}$		& 2438911.5683$^{+7}_{-14}$	& $-42.71^{+66}_{-48}$			& $-0.1560^{+24}_{-17}$	& 83.0$^{+2.3}_{-2.9}$		& 2422100$^{+1200}_{-1000}$	& 0.811$^{+50}_{-82}$	& 134$^{+15}_{-8}$	& 0.00787$^{+93}_{-81}$	& 1.36$^{+16}_{-14}$	& 0.00037$^{+13}_{-7}$		& 0.0540$^{+63}_{-40}$		& 0.84$^{+13}_{-13}$	& 1.095(86)		& 279\\
\vspace*{0.15cm} SS Leo 	& 1	& 0.62634133$^{+25}_{-30}$		& 2441781.3914$^{+17}_{-19}$	& --					& --			& 110.7$^{+7.3}_{-5.2}$		& 2453240$^{+440}_{-530}$	& 0.475$^{+51}_{-28}$	& 141$^{+10}_{-12}$	& 0.0517$^{+38}_{-30}$	& 8.95$^{+66}_{-51}$	& 0.0585$^{+63}_{-58}$		& 0.384$^{+17}_{-20}$		& 2.737$^{+68}_{-45}$	& 1.09(13)		& 128\\
\vspace*{0.15cm} AV Peg 	& 2	& 0.390375282$^{+22}_{-22}$		& 2441552.28877$^{+76}_{-34}$	& $+45.96^{+22}_{-42}$ 			& $+0.1679^{+8}_{-15}$	& 47.7$^{+1.8}_{-2.2}$		& 2455880$^{+280}_{-460}$	& 0.44$^{+18}_{-9}$	& 86$^{+18}_{-16}$	& 0.00591$^{+52}_{-10}$	& 1.024$^{+90}_{-17}$	& 0.00047$^{+15}_{-3}$		& 0.0589$^{+61}_{-16}$		& 0.71$^{+18}_{-4}$	& 1.010(63)		& 518\\
\vspace*{0.15cm} AT Ser 	& 1	& 0.74656003$^{+17}_{-18}$		& 2436093.2609$^{+26}_{-49}$	& --					& --			& 85.6$^{+1.2}_{-1.4}$		& 2449440$^{+550}_{-310}$	& 0.456$^{+47}_{-50}$	& 217.7$^{+7.9}_{-4.6}$	& 0.1157$^{+35}_{-75}$	& 20.0$^{+0.6}_{-1.3}$	& 1.10$^{+10}_{-20}$		& 1.90$^{+14}_{-22}$		& 7.83$^{+28}_{-54}$	& 1.12(16)		& 84\\
\vspace*{0.15cm} RV UMa$^{B}$	& 2	& 0.468062850$^{+34}_{-32}$		& 2438459.9318$^{+16}_{-17}$	& $+4.70\,^{+62}_{-54}$			& $+0.0172^{+23}_{-20}$	& 66.93$^{+84}_{-78}$		& 2449520$^{+510}_{-570}$	& 0.404$^{+55}_{-49}$	& 317$^{+10}_{-11}$	& 0.02657$^{+90}_{-81}$	& 4.60$^{+16}_{-14}$	& 0.0217$^{+24}_{-21}$		& 0.251$^{+11}_{-9}$		& 2.24$^{+11}_{-9}$	& 1.105(78)		& 336\\
\vspace*{0.15cm} BB Vir$^{B?}$	& 2	& 0.47109958$^{+6}_{-21}$		& 2436232.2426$^{+48}_{-29}$	& $+36.3\,^{+1.5}_{-0.3}$		& $+0.1325^{+54}_{-9}$	& 92.7$^{+2.7}_{-8.8}$		& 2446900$^{+6000}_{-1100}$	& 0.24$^{+18}_{-1}$ 	& 296$^{+76}_{-10}$	& 0.0248$^{+12}_{-41}$	& 4.30$^{+21}_{-70}$	& 0.0092$^{+29}_{-32}$		& 0.177$^{+15}_{-32}$		& 1.42$^{+23}_{-11}$	& 1.11(14)		& 109\\
\hline
\end{tabular}\label{Tab:LiTEtable}
}

{\scriptsize {\bf Notes.} $^{(*)}$ Columns contain following parameters: Star -- name of the star in the GCVS, Model -- type of calculated model (1 = only \lt, 2 = \lt~+~parabola), $P_{\rm puls}$ -- pulsation period, $M_{0}$ -- zero epoch of pulsations, $\dot{P}_{\rm puls} = \beta$ -- relative rate of changes of pulsation period, $P_{\rm orbit}$ -- orbital period, $T_{0}$ -- time of periastron passage, $e$ -- numerical eccentricity, $\omega$ -- argument of periastron, $A$ -- $a_{1}\sin i$ in light days (semi-amplitude of \lt~$A_{\rm LiTE}$ can be calculated as $A_{\rm LiTE} = A\,\sqrt{1-e^{2}\,\cos^{2}\omega}$), $a_{1}\sin i$ -- projection of semi-major axis of primary component $a_{1}$ according to the inclination of the orbit $i$, $f(\mathfrak{M})$ -- mass function, $\mathfrak{M}_{\rm 2, min}$ -- the lowest mass of the second component, the value was calculated for inclination angle $i=90^{\circ}$ and adopted mass of primary $\mathfrak{M}_{1} = 0.6$\,$\mathfrak{M}_{\odot}$, $K_{1}$ -- semi-amplitude of RV changes primary component (RR Lyrae star), $\chi_{\rm R}^2$ -- normalised value of $\chi^{2}$, where $\chi^{2}_{\rm R} = \chi^{2}/(N_{\rm max}-g)$ for number of used measurements $N_{\rm max}$ and number of free (fitted) parameters $g$ (only \lt~$g=7$, \lt~+~parabola $g=8$), $N_{\rm max}$ -- number of used maxima timings, $^{(B)}$ -- \Blazhko~effect is known in the star, more information in the text, $^{(B?)}$ -- the star is suspected from \Blazhko~effect, more information in the text.} 
\end{minipage}
\end{table*}

New photometric measurements in \textit{BVRI} bands were performed for VX Her at Masaryk University Observatory (MUO), Brno, Czech Republic during 13 nights (April\,--\,August 2014) with a 24-inch (0.6-m) Newtonian telescope and a G2-0402 CCD camera. Other measurements in phases near maximum brightness for the stars SS~Leo and AT~Ser were also obtained in Brno with a small 1-inch telescope in the $green$-band \citep[description of the telescope in][]{liska2014}.

All the maxima used from the GEOS database, with corresponding references, maxima determined from sky surveys measurements, and from our observations, are available via CDS. For each star the interval covered by these measurements is given in Table~\ref{Tab:NumberMaximatable}.

%*************************************************************************
\section{Results on stars that are suspect}\label{resultssec} 
%*************************************************************************
The process of \lt~fitting described in \citet{liska2015} was applied on our selected target stars whose identification and basic light ephemeris are in Table~\ref{Tab:BasicCharTable}. Each of these stars shows significant period changes. Several of them were previously reported as possible members of binary stars, but with undetermined orbital periods, or only very roughly estimated periods.

Some of the \oc~changes can be interpreted directly as large amplitude cyclic changes. Such dependences were fitted as a whole and are described by our model 1. In several cases the orbital period is similar to the length of the time-base. Therefore model~1 should be considered only as a possible solution, in which the observed changes may not be cyclic at all. Some of the stars' \oc~diagrams can be interpreted as a low amplitude cyclic change superimposed on a secular period change. This situation is described by our model 2. However, in some cases both interpretations could be possible and, therefore, both models were computed. Our results with the parameters describing the proposed orbits are in Table~\ref{Tab:LiTEtable}. The columns are described in the notes below the table. Except for orbital parameters this table also contains the period-change rate (if present), goodness of fit indicator ($\chi^{2}_{\rm R}$) and the number of used maxima timings. For calculation of minimum masses of components we assumed an RR Lyrae mass of 0.6\,\M. 

In agreement with expectations, it seems that companions of the majority of our sample stars are low-mass stars. However, in~RZ~Cet the companion is probably a white dwarf or neutron star, and in AT~Ser it is probably a black hole. The studied stars are discussed in~detail in the next sections.

Available literature contains a relatively rich sample of original RV measurements (Table \ref{Tab:RVPublicationtable}) which could possibly be used for verification of our results, because our models allow the computation of systemic variations in RV \citep[for details see][]{liska2015}. Unfortunately they are in most cases of an insufficient number, time span, or quality for such usage. Systemic velocities for our targets from \citet{layden1994}, \citet{fernley1997} or \citet{solano1997}, which are typically based on 2 or 3 unpublished measurements, are of insufficient quality. During our study we found that among our targets some of the RV measurements may not be reliable -- mainly old values from \citet{abt1970,abt1973} can differ by more than 30\,km\,s$^{-1}$ compared to other datasets\footnote{\citet{layden1994} analysed a large sample of mean RVs from \citet{payne-gaposchkin1954} for RR~Lyraes and estimated the uncertainty of these measurements as 35\,km\,s$^{-1}$. In many cases original values for these mean RV measurements were published later in \citet{abt1970,abt1973}.}. This shows the need for new accurate spectroscopy.

\begin{table}
\begin{center}
\caption{Sources of radial velocity measurements for individual RR Lyrae stars.}
{%\tiny
\def\arraystretch{1.5}
\tabcolsep=1.3pt
\begin{tabular}{llp{1.0cm}llp{0.15cm}}
\hline
Star	&	\hspace*{0.3cm}Ref. & & Star	&	\hspace*{0.3cm}Ref. &\\
\hline
RS Boo	&	\hspace*{0.3cm}2, 8, 14 					&& SS Leo	&	\hspace*{0.3cm}1, 3, 5, 7 &\\
RU CVn	&	\hspace*{0.3cm}2 							&& AV Peg	&	\hspace*{0.3cm}2, 9, 14 &\\
RZ Cet	&	\hspace*{0.3cm}4 							&& AT Ser	&	\hspace*{0.3cm}2, 6, 7 &\\
S Com	&	\hspace*{0.3cm}2, 7 						&& RV UMa	&	\hspace*{0.3cm}2, 6, 12 &\\
SU Dra	&	\hspace*{0.3cm}1, 9, 10, 11, 13 			&& BB Vir	&	\hspace*{0.3cm}2&\\
VX Her	&	\hspace*{0.3cm}2, 14 & & &\\
\hline
\end{tabular}}	\label{Tab:RVPublicationtable}
\end{center}
\vspace*{0.2cm}
1\,--\,\citet{abt1970},	%Abt 1970
2\,--\,\citet{abt1973}, %Abt 1973
3\,--\,\citet{carrillo1995}, %Carrillo et al. 1995
4\,--\,\citet{colacevich1950}, %Colacevich 1950
5\,--\,\citet{fernley1990}, %Fernley et al. 1990
6\,--\,\citet{fernley1993b}, %Fernley, Skillen \& Burki 1993
7\,--\,\citet{hawley1985}, %Hawley \& Barnes 1985
8\,--\,\citet{jones1988}, %Jones, Carney \& Latham 1988
9\,--\,\citet{liu1989}, %Liu \& Janes 1989
10\,--\,\citet{oke1962}, %Oke et al. 1962
11\,--\,\citet{preston1965}, %Preston 1965
12\,--\,\citet{preston1967}, %Preston \& Spirad 1967
13\,--\,\citet{varsavsky1960}, %Varsavsky 1960
14\,--\,\citet{woolley1966}. %Woolley \& Aly 1966
\end{table}

% ===========================================================
% ===========================================================
\subsection{RS Boo}\label{resultsrsboo}

RS Boo is one of the best studied RR Lyraes. Except for long-term period changes, it shows modulation (\Blazhko~effect) with a period of about 533\,d \citep{oosterhoff1946} and a possible shorter period of between 58\,--\,62\,d \citep{kanyo1980}. More recent analyses showed multiple \Blazhko~modulation with periods of 532.481\,d \citep{leborgne2012} and 41.3\,d and 62.5\,d \citep{skarka2014}. The idea that RS~Boo could be a binary comes from \citet{kanyo1980} who investigated its \oc~diagram. After subtraction of a parabola he proposed an orbital period of about 70\,years based on residuals. \citet{kanyo1986} also found that the brightening during the 62-day cycle is lower in blue than in yellow, contrary to the long 530-day cycle. He guess that it could be due to a cooler companion.

\begin{figure}%[htbp]
\centering
\includegraphics[width=0.95\hsize]{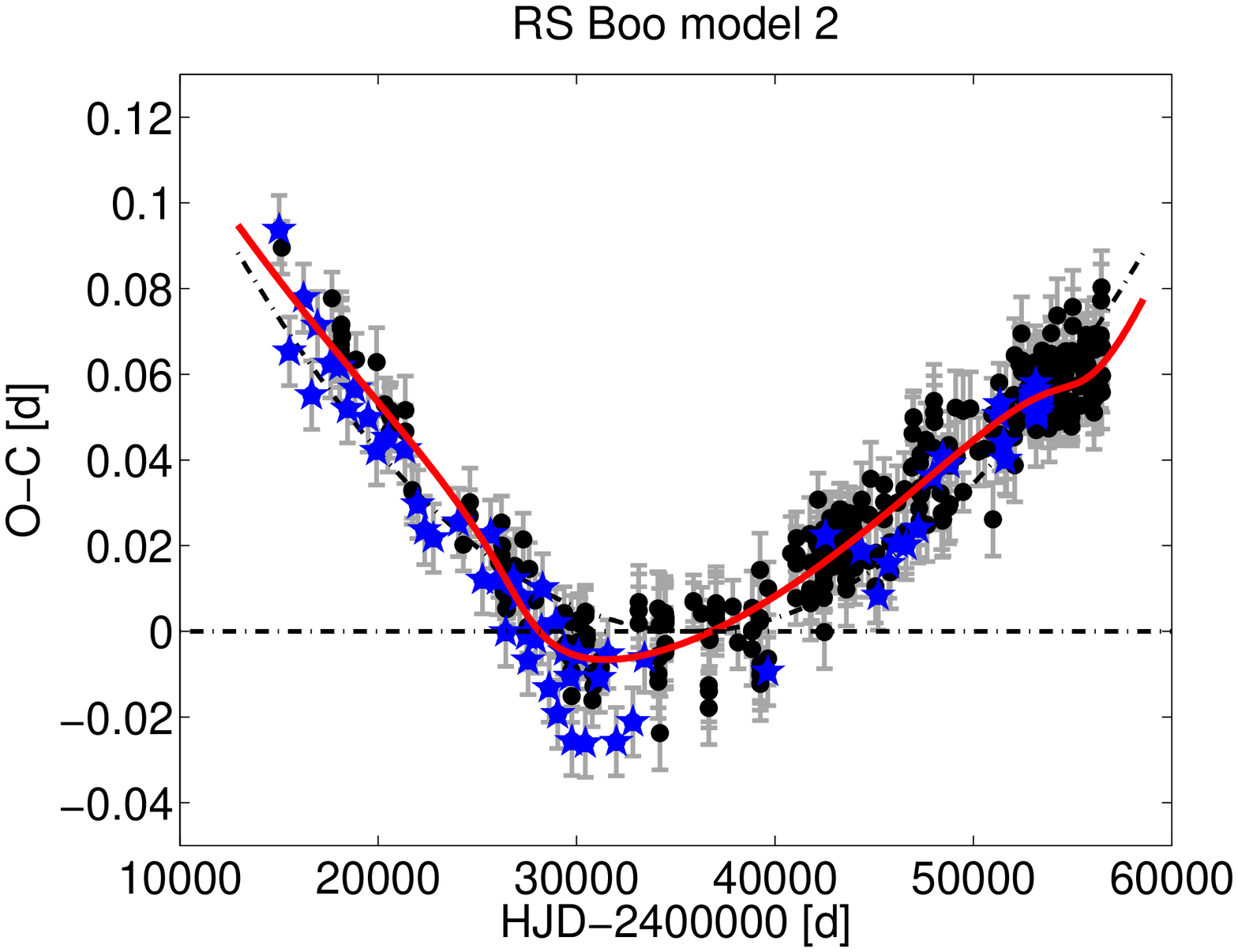}
\includegraphics[width=0.95\hsize]{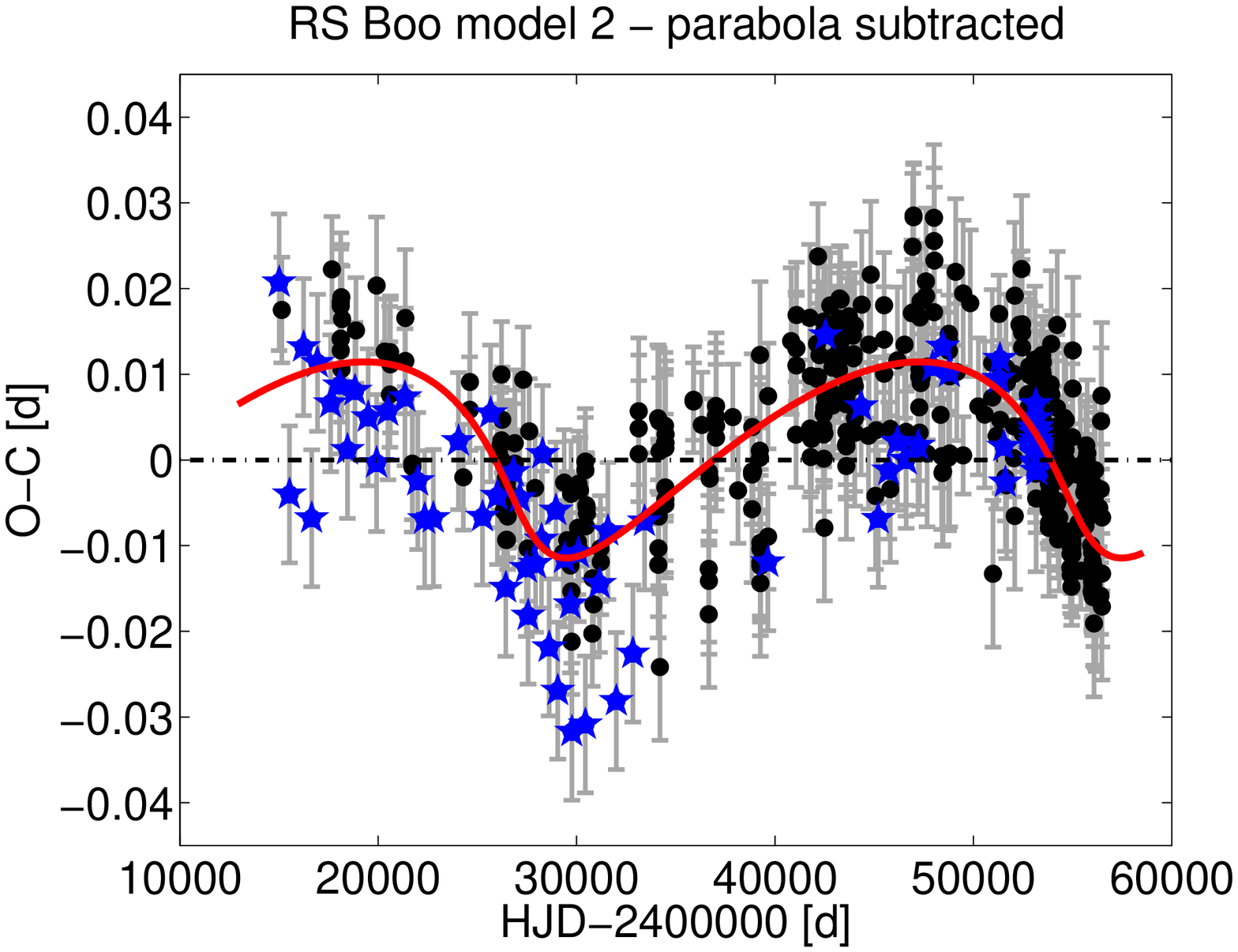}
\caption{\oc~diagram of RS Boo together with our model 2 (solid line, top panel) and variation after subtracting the parabolic trend (bottom panel). Black circles and blue stars display maxima adopted from the GEOS database and new maxima determined in this work (from sky surveys measurements and from our observations), respectively.}
\label{Fig:RSBoo}
\end{figure}

The shape of the \oc~diagram of RS Boo (significantly extended with values based on the DASCH project) is apparently asymmetric (Fig. \ref{Fig:RSBoo}) allowing direct interpretation as a consequence of \lt~(our model 1) and as a secular period lengthening with superimposed \lt~(model 2, Fig. \ref{Fig:RSBoo}). Model 1 gives an orbital period of 147\,yr and a minimum mass of the companion of 0.264\,$\mathfrak{M_{\odot}}$, while model 2 gives an orbital period of 77\,yr and a minimum mass of the second component of 0.097\,$\mathfrak{M_{\odot}}$. Although the shape of the \oc~also suggests non-symmetrical period lengthening\footnote{\citet{leborgne2007} performed a third-order least-squares fit.}, the indication that the dependence should go makes this explanation less plausible. Because both models have the same quality of the fit ($\chi^{2}_{\rm R}=1.082$) and both secondary component masses are possible, it is very difficult to choose the correct solution. Only further precise photometric and spectroscopic measurements could help.

\begin{figure}%[htbp]
\centering
\includegraphics[width=0.95\hsize]{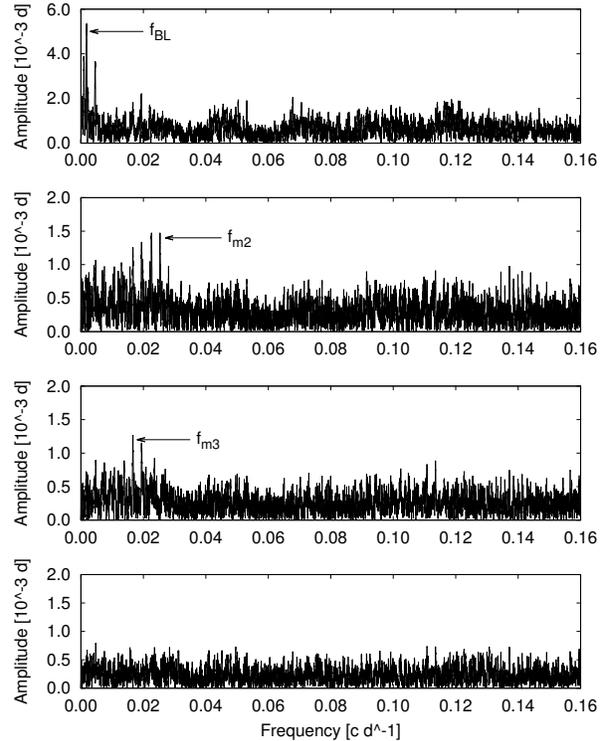}
\caption{Frequency spectra of photoelectric and CCD maxima for RS Boo. Steps in prewhitening with labelled frequencies go from the top to the bottom panel, where residuals are shown.}
\label{Fig:RSBooResiduals}
\end{figure}

\begin{figure}%[htbp]
\centering
\includegraphics[width=0.95\hsize]{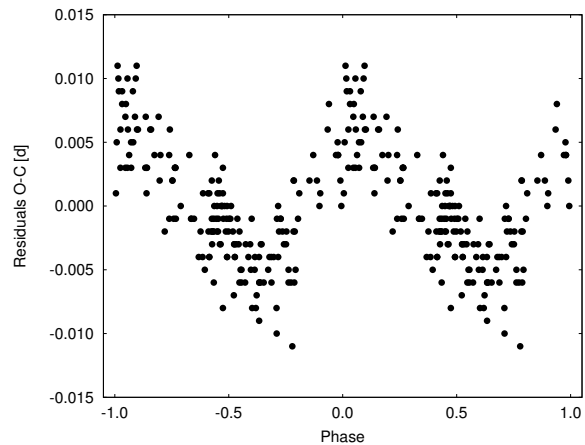}
\caption{Residual \oc~values of RS Boo after subtracting model 1 phased with the found \Blazhko~period of 550.5\,d. Vertical scatter is caused by additional modulation components.}
\label{Fig:RSBooPhased}
\end{figure}

After subtraction of both models we searched for additional periodicity in residuals. Both models give almost the same residuals, therefore the same periods were found in both cases. The most significant peak in the frequency spectrum (top panel of the Fig. \ref{Fig:RSBooResiduals}) relates to a \Blazhko~period of $P_{\mathrm{BL}}=535(4)$\,d (for all available data), which is within the error margin the same period as from previous studies \citep[e.g.][]{kanyo1986,leborgne2007,leborgne2012}, and $P_{\mathrm{BL}}=550.5(1.0)$\,d when only more recent photoelectric and CCD measurements were used (data phased with this period are in Fig.~\ref{Fig:RSBooPhased}). After prewhitening with this peak, no other significant peak was revealed taking all measurements into account. On the other hand, when frequency spectra of CCD and photoelectric data were investigated, three additional peaks were identified. The first of them, with unclear interpretation, corresponds to a period of 2400\,d. The other two peaks, corresponding to periods $P_{\mathrm{m2}}=39.5(7)$\,d and $P_{\mathrm{m3}}=60.0(1)$\,d (two middle panels in Fig.~\ref{Fig:RSBooResiduals}), correlate well with the secondary modulation periods from literature. Therefore we argue that the secondary modulation components are real and need to be refined by additional observations.

Our first model of \lt~suggests RV variations with a semi-amplitude of $K_{1}\sim 1.7$\,km\,s$^{-1}$ (or only 1\,km\,s$^{-1}$ from model~2) which is too small for reliable confirmation with the available RV measurements (see Table~\ref{Tab:RVPublicationtable}). It is because the only measurements with a sufficient accuracy, the ones of \citet{jones1988}, span only 108 days in various phases of the 550-day cycle, and therefore they are useless for our purposes. At least these data allowed us to reject the possibility that the 550-day cycle could be a consequence of an additional \lt, because such \lt~would produce periodic variations in the systemic velocity with a semi-amplitude of 20.9\,km\,s$^{-1}$ which, after subtraction from the values adopted from \citet{jones1988}, results in an increase in the scatter of the phased RV curve.

% ===========================================================
% ===========================================================
\subsection{RU CVn}\label{resultsrucvn}

\begin{figure}
\centering
\includegraphics[width=0.95\hsize]{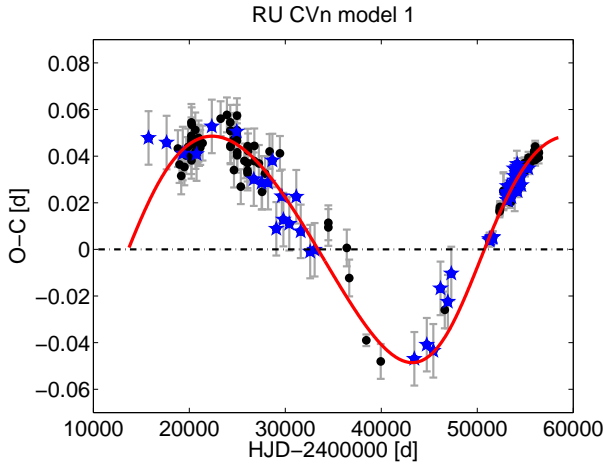}
\caption{\oc~diagram of RU CVn together with our model~1. Symbols are the same as in Fig.~\ref{Fig:RSBoo}.}
\label{Fig:RUCVn}
\end{figure}

Possible long-term cyclic changes in the pulsation period of probably 60000\,yr\footnote{This value should probably be in days not in years.} was discussed by \citet{husar2003}, and due to the complicated shape of the \oc~diagram was mentioned in \citet{leborgne2007}. The dependence in Fig.~\ref{Fig:RUCVn} clearly shows a cyclic variation. Thus only model 1 was applied, which gives an orbital period of about 102\,yr and a minimum mass of the companion of 0.39\,\M. However, orbital parameters based on our model should be considered only as preliminary, because only one cycle has been completed up to now.

RU CVn has only 2 single RV measurements available in literature \citep{abt1973}. Both values were obtained with a nine-year difference at nearly the same pulsation phase of 0.33, but they differ significantly (+21.7, $-$58.9)\,km\,s$^{-1}$. The big difference in RVs cannot be explained by the difference in systemic velocity based on our \lt~model, because changes due to two different phases of the orbit should only be about 1\,km\,s$^{-1}$. In addition, the semi-amplitude of RV changes $K_{1}$ is only 2.6\,km\,s$^{-1}$. One of the values is probably incorrect or values are highly scattered (this is a more plausible explanation), see Sect.~\ref{resultssec}.

% ===========================================================
% ===========================================================
\subsection{RZ Cet}\label{resultsrzcet}
\begin{figure}
\centering
\includegraphics[width=0.95\hsize]{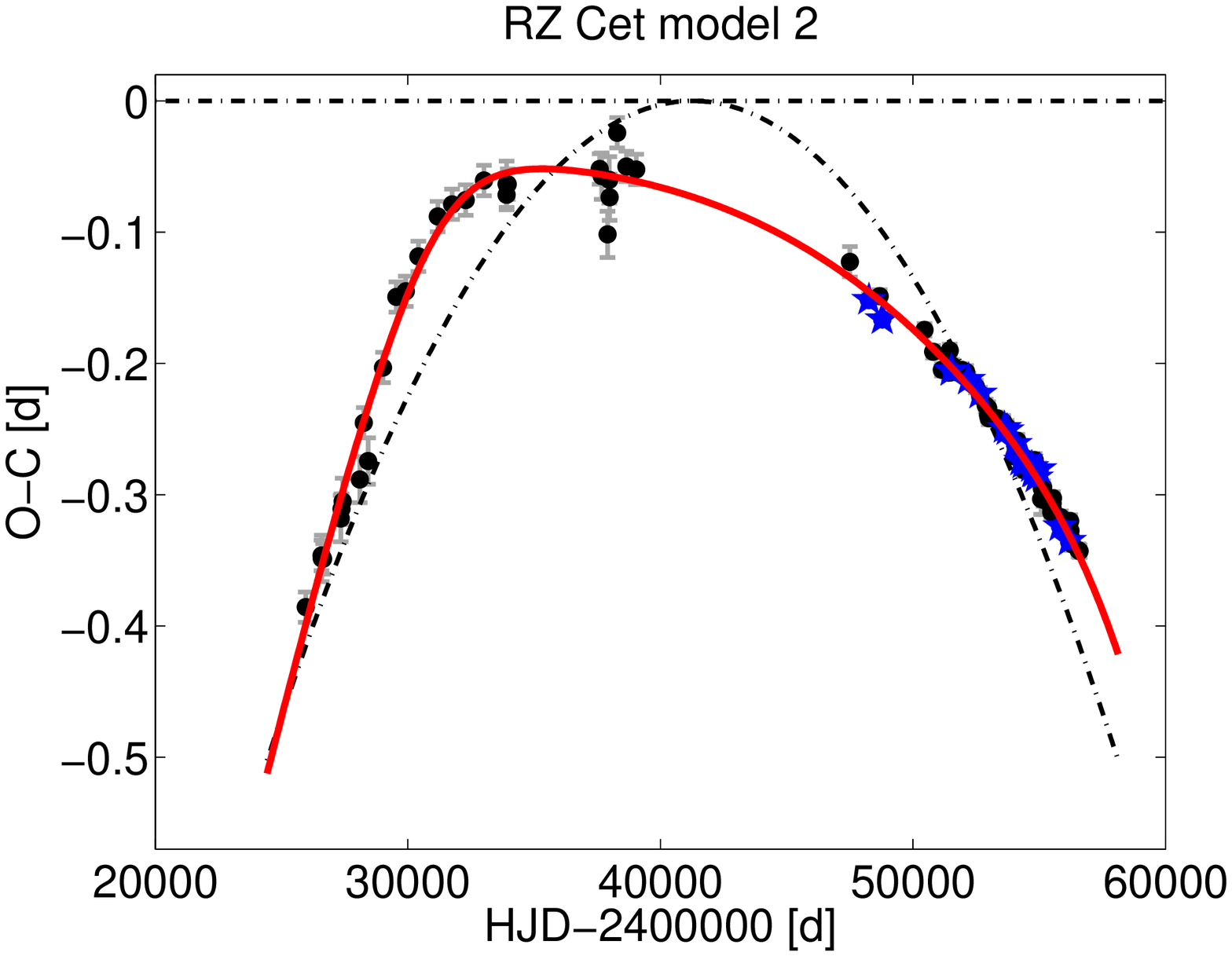}
\includegraphics[width=0.95\hsize]{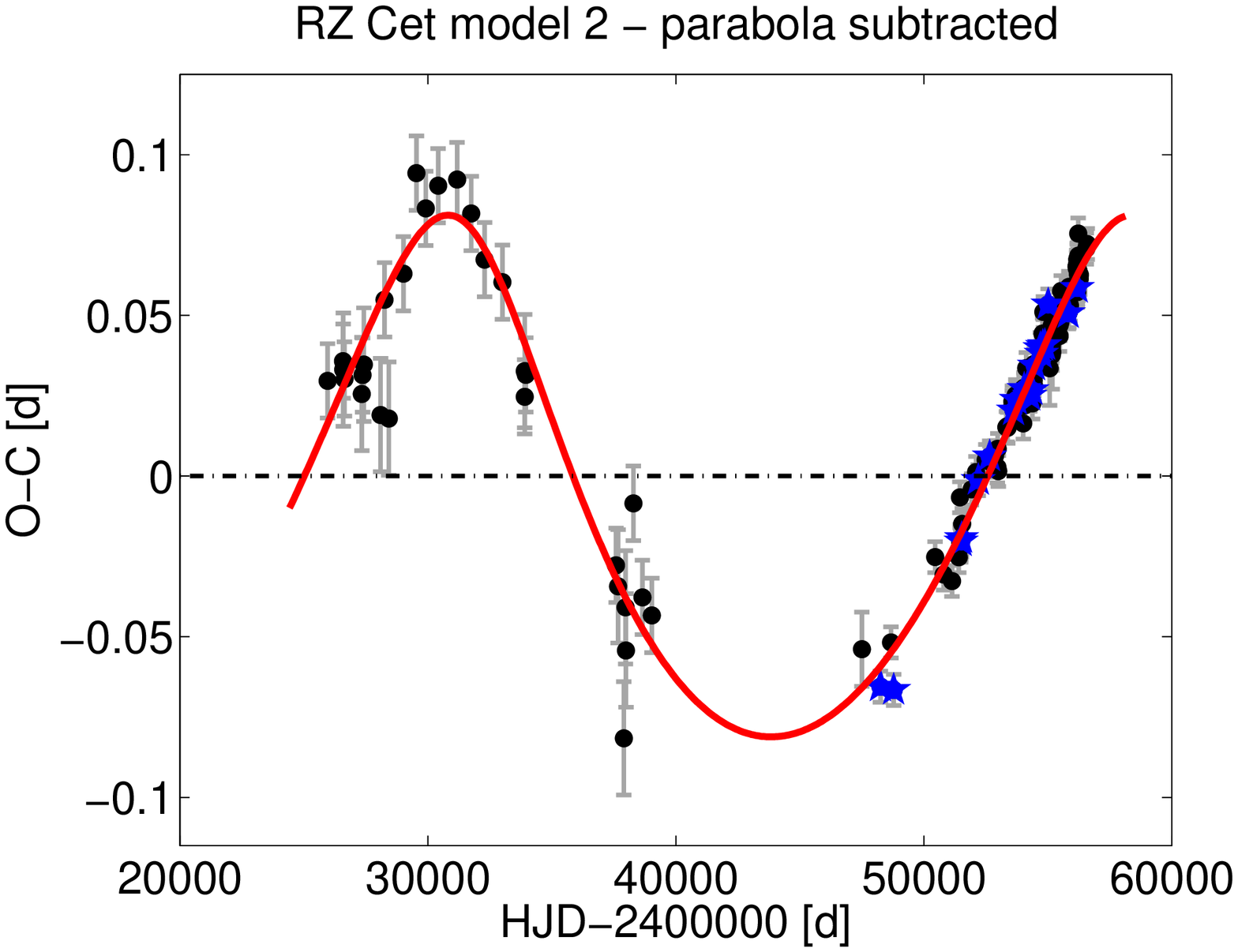}
\caption{\oc~diagram of RZ Cet together with model 2 (top panel) and variation after subtracting the parabolic trend (bottom panel). Symbols are the same as in Fig.~\ref{Fig:RSBoo}.}
\label{Fig:RZCet}
\end{figure}

Period variations of RZ Cet in its \oc~diagram are well pronounced (Fig.~\ref{Fig:RZCet}) and were studied e.g. by \citet{leborgne2007}. They fitted the \oc~dependence with a parabola ($\dot{P}_{\rm puls}=-157.9(8.8)\times 10^{-11}$\,d\,d$^{-1}$) and subsequently found oscillations in residuals with a period of 12500\,d superimposed on the parabolic trend. They mentioned that this oscillation could be the consequence of \lt. 

We tested the proposed scenario using a similar approach -- simultaneous fitting of a parabola and \lt~and found $\dot{P}_{\rm puls}=-180.8^{+7.6}_{-6.9}\times 10^{-11}$\,d\,d$^{-1}$, and a \lt~period of 75\,yr (27600\,d) which is more than two times more than the value from \citet{leborgne2007}. Nevertheless, the time-baseline has a similar value (84\,yr) and thus the orbital period and other parameters (such as companion mass $\mathfrak{M}_{\rm 2,min} \sim 1.15$ \,\M) are very preliminary.

However, variations in the \oc~diagram allow for another, less probable scenario: that the \oc~dependence itself could be interpreted as a \lt~ (without parabola, model 1) with incomplete cycle. The insufficient time span caused the modelling to be unstable and the \lt~model parameters had to be manually adjusted. This 'solution' gives an orbital period of at least 86 years with an amplitude suggesting a high-mass companion with a minimum mass of 7.5\,\M. According to the discussion in Sect.~\ref{introductionsec} this should be a black hole. Since the proposed orbital period is more than 80 years, the system would be well detached and therefore no mass transfer would be possible, and the black hole would manifest itself only via gravitational effects on the RR Lyrae component. Although the presence of a black hole could be surprising, RZ Cet would not be the first object among RR Lyrae type stars with a possible black hole (see the discussion about BE~Dor in Sect.~\ref{rrlitesec}). 

\citet{kovacs2005} identified RZ Cet as a modulated star, but the period of the proposed \Blazhko~effect is not known. \citet{skarka2014} analysed data of RZ Cet from the ASAS survey and found no indications of modulation. It is possible that high-amplitude \oc~changes could be attributed to the \Blazhko~effect with an extremely long period, but such a long modulation period has not been found yet (see Sec~\ref{rrocblazhkosubsec}). Analysis of the residuals also did not show any signs of modulation.

Only one single RV measurement for RZ Cet is available \citep{colacevich1950}, which is inappropriate for our purposes. 
Spectroscopic measurements are definitely highly needed to confirm the binarity of RZ Cet.

% ===========================================================
% ===========================================================
\subsection{S Com}\label{resultsscom}

\begin{figure}
\centering
\includegraphics[width=0.95\hsize]{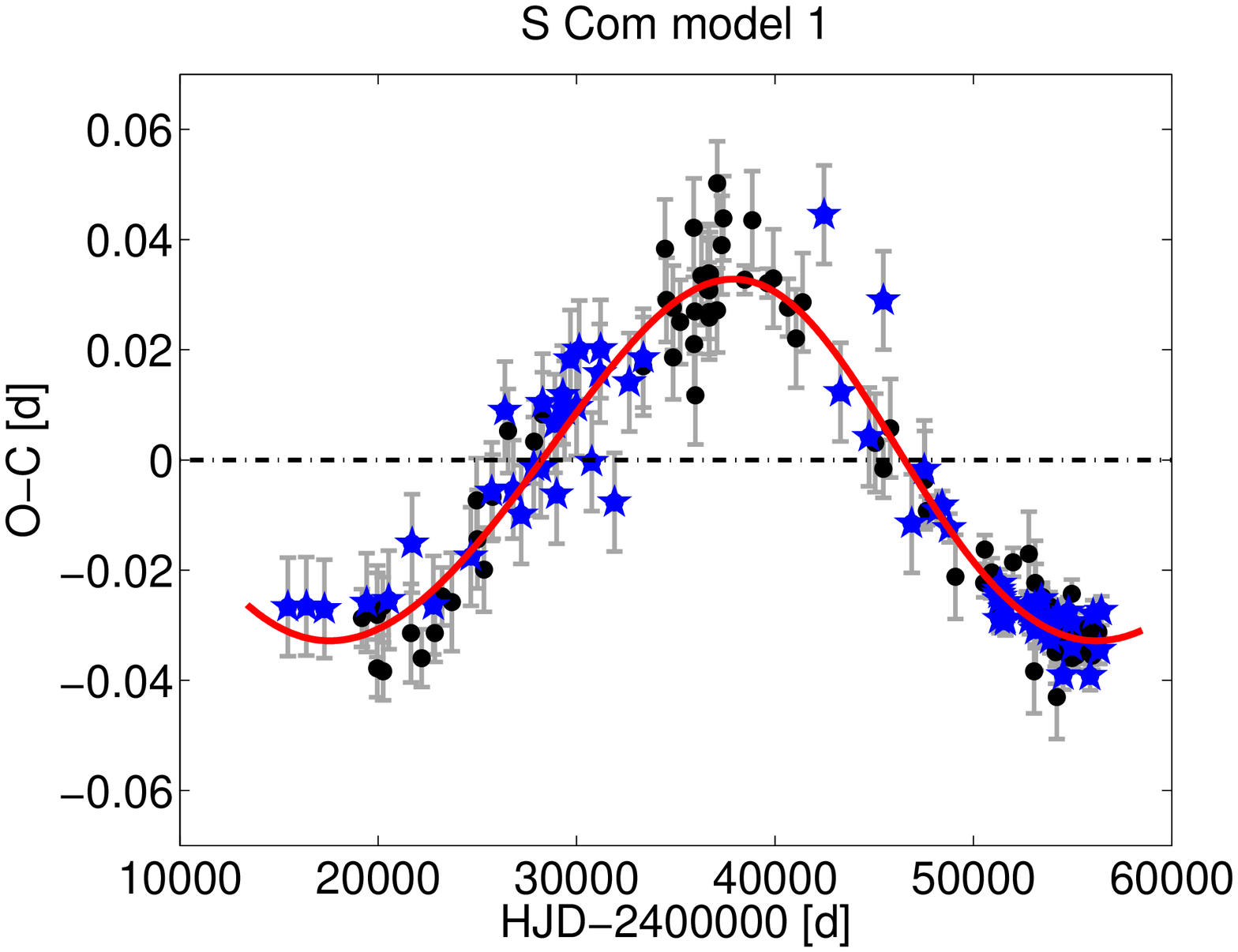}
\includegraphics[width=0.95\hsize]{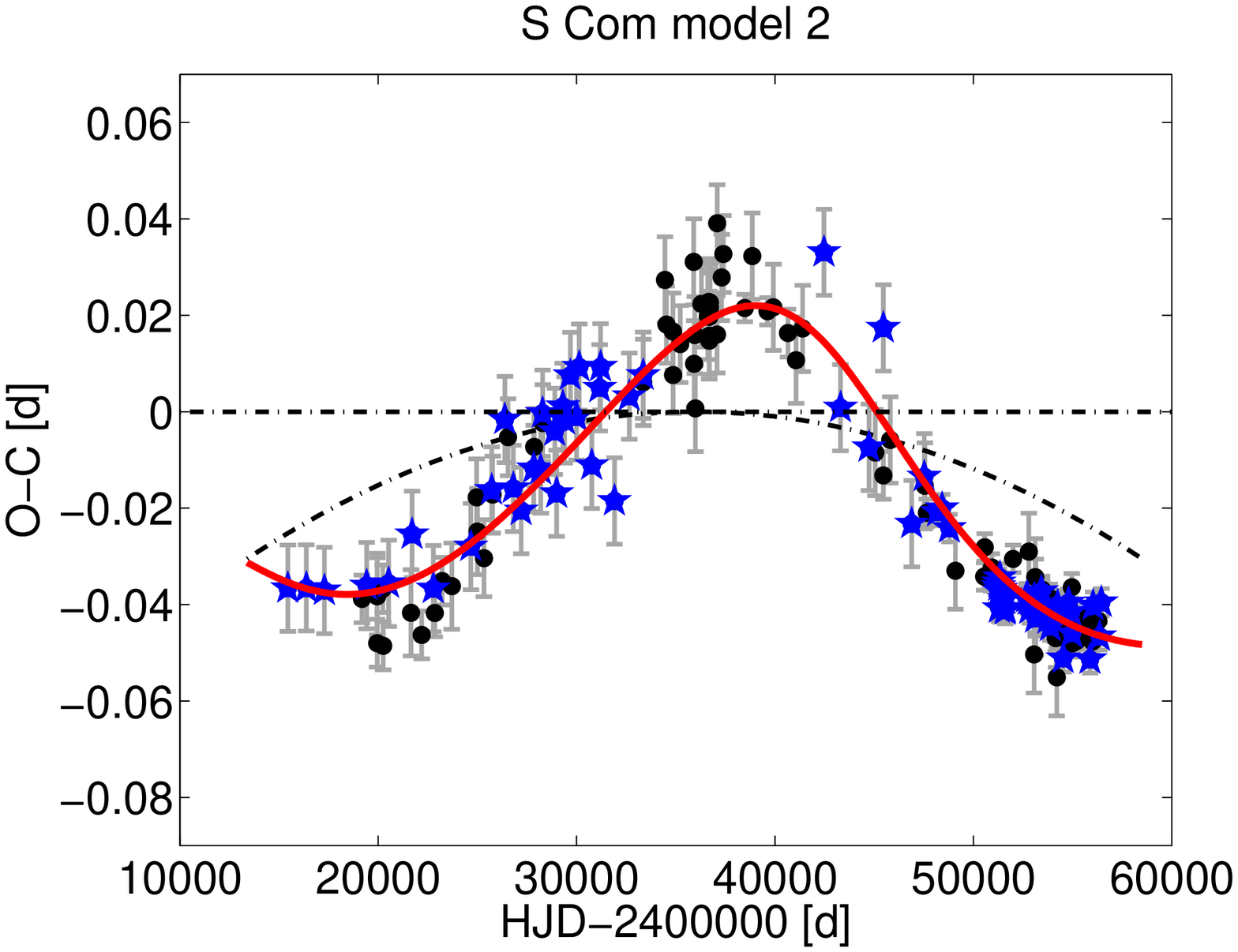}
\includegraphics[width=0.95\hsize]{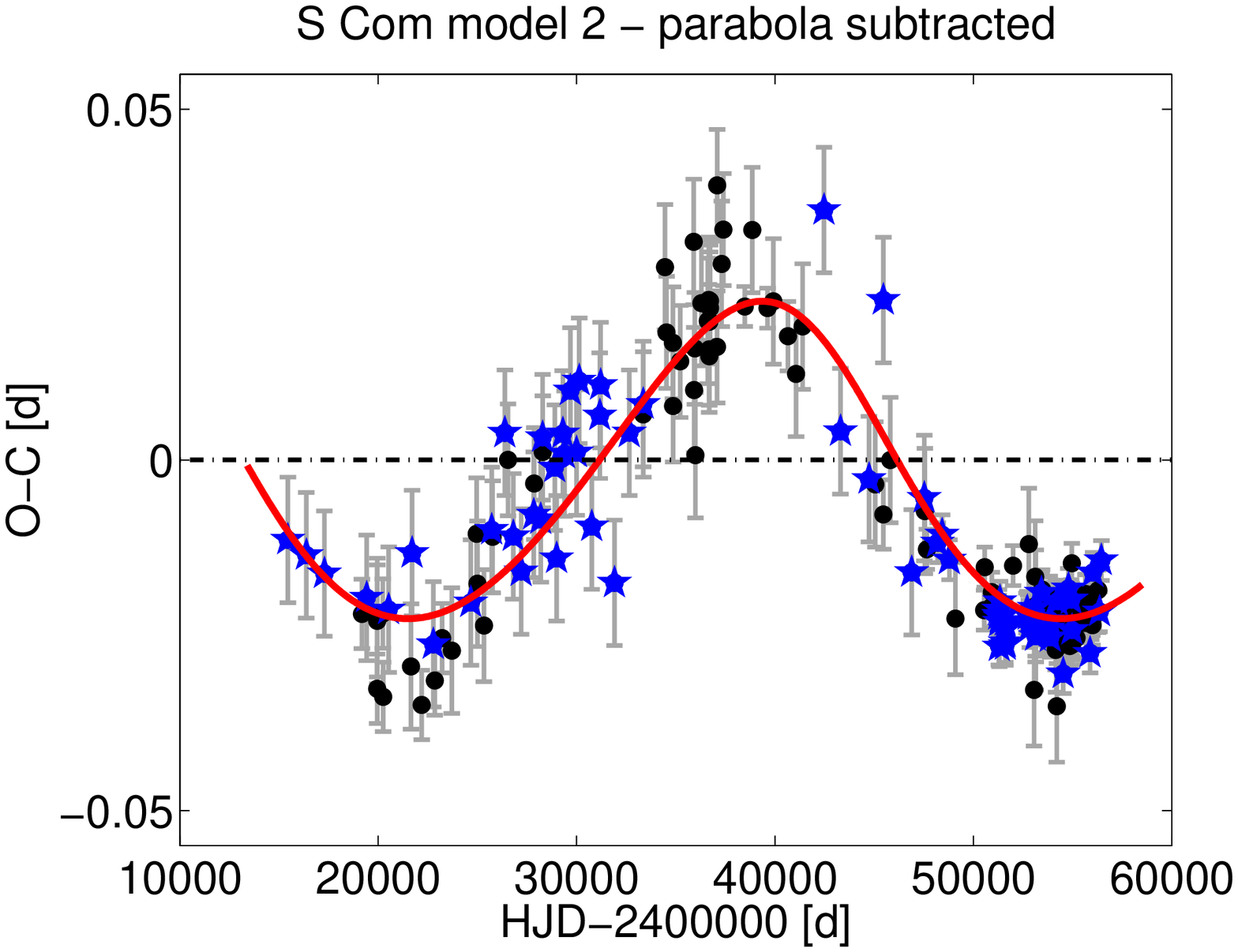}
\caption{\oc~diagram of S Com together with our model 1 (top panel), model 2 (middle panel) and variation after subtracting the parabolic trend (bottom panel). Symbols are the same as in Fig.~\ref{Fig:RSBoo}.}
\label{Fig:SCom}
\end{figure}

S Com is another promising candidate for a binary system. The information about possible cyclic changes with a long period was mentioned without closer details in \citet{leborgne2007}. Probable \lt~with a period of about 106\,yr is well demonstrated in our \oc~diagram (model~1, the top panel in the Fig.~\ref{Fig:SCom}). Currently, the pulsation period is at its the shortest, but in the next few years it should start rising again. 

Our second model of \lt~of S Com including the parabolic trend (the middle panel in Fig.~\ref{Fig:SCom}) gives a shorter orbital period of 90\,yr and also lower orbital parameters such as the projection of the semi-major axis or the limiting mass of the secondary component. The rate of period changes $\dot{P}_{\rm puls}=-7.0^{+1.2}_{-4.1}\times 10^{-11}$\,d\,d$^{-1}$ significantly differs from the value $-26.4(1.1)\times 10^{-11}$\,d\,d$^{-1}$ \citep{leborgne2007} mainly due to inclusion of \lt~in the model.

There are two RV datasets available in the literature (Table~\ref{Tab:RVPublicationtable}). However, expected changes of the systemic velocity caused by binarity with a semi-amplitude of only 1.6\,km\,s$^{-1}$ (model~1) or 1.3\,km\,s$^{-1}$ (model~2) are completely saturated by the RV changes caused by pulsations. 

% ===========================================================
% ===========================================================
\subsection{SU Dra}\label{resultssudra}
\begin{figure}
\centering
\includegraphics[width=0.95\hsize]{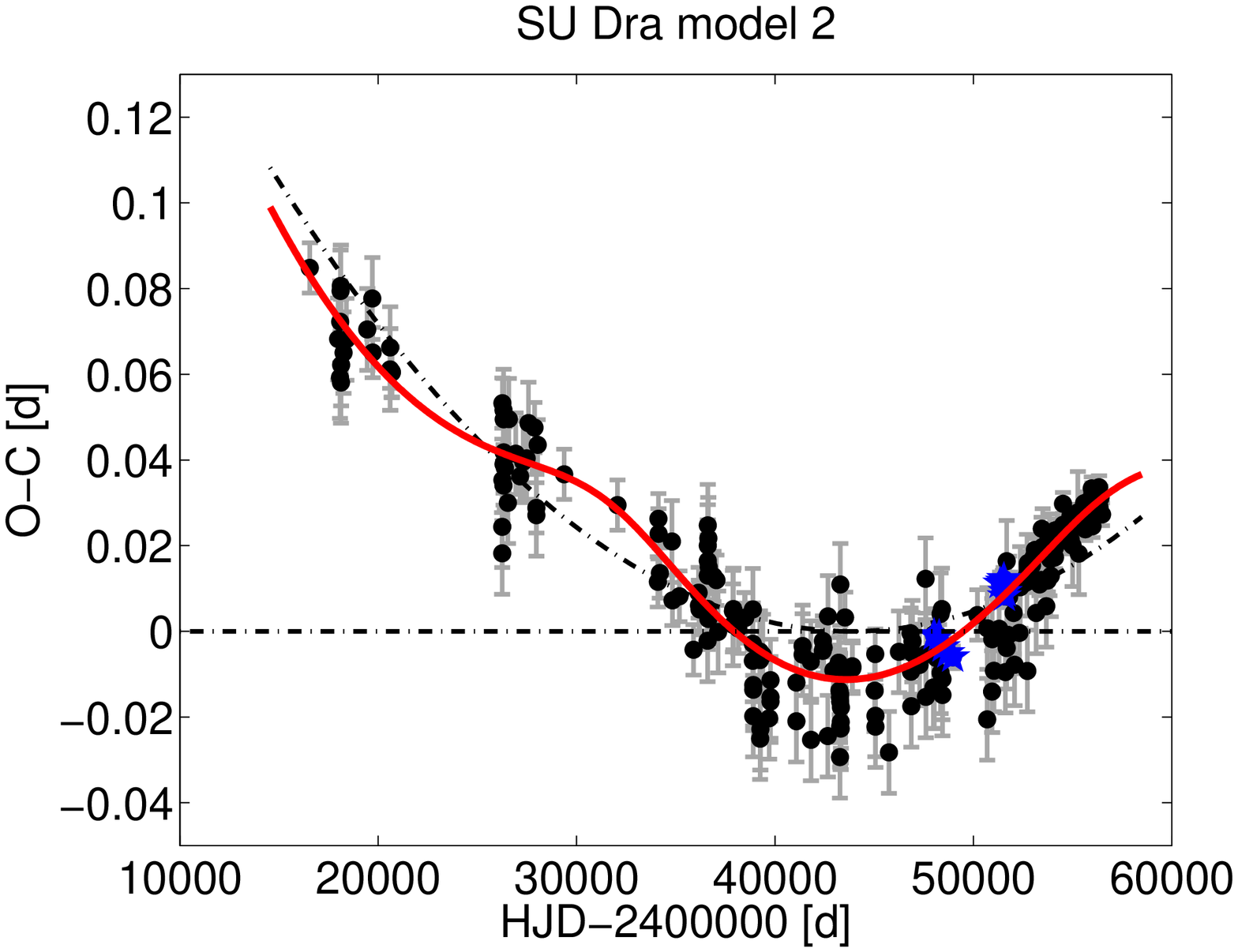}
\includegraphics[width=0.95\hsize]{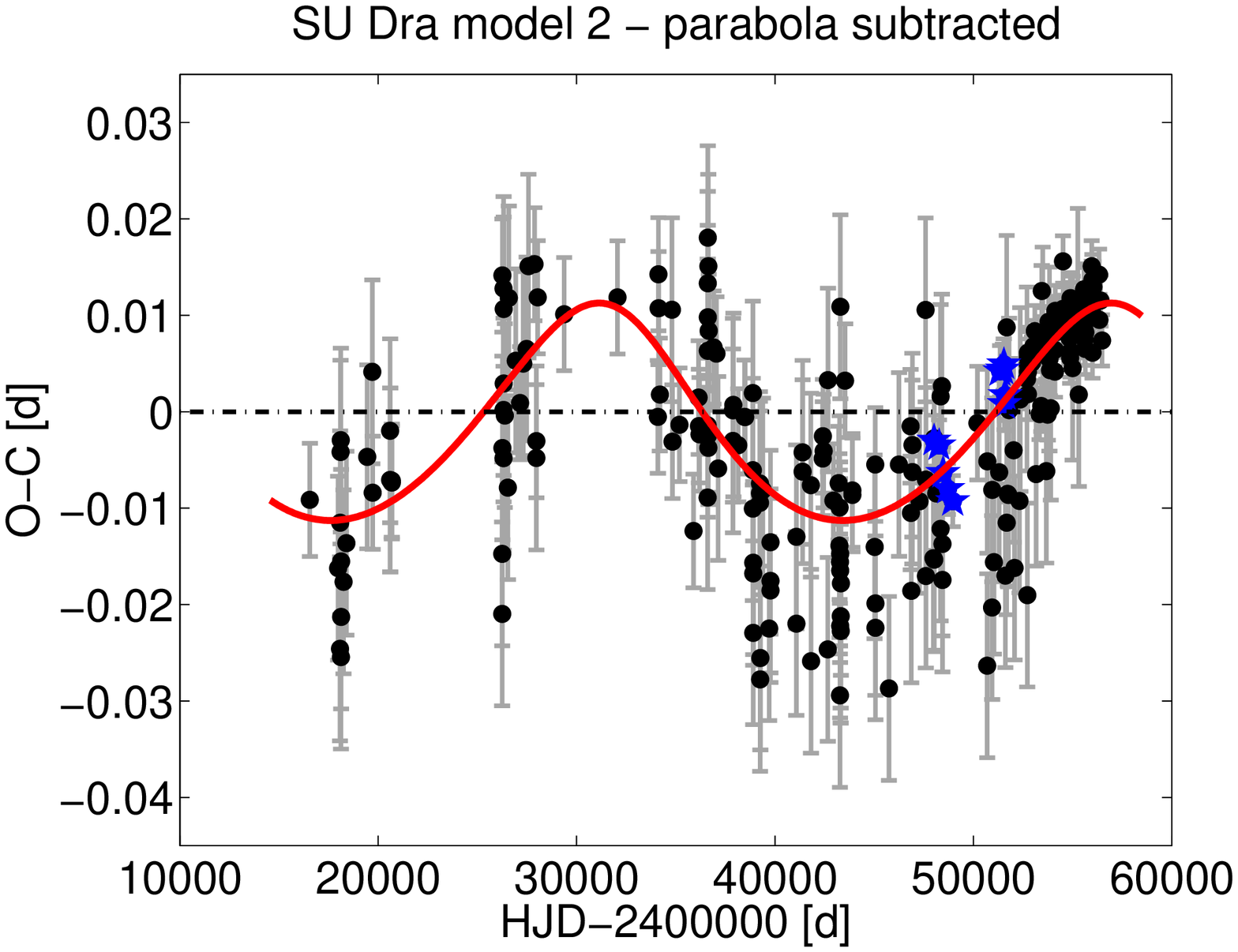}
\caption{\oc~diagram of SU Dra together with our model 2 (top panel) and variation after subtracting the parabolic trend (bottom panel). Symbols are the same as in Fig.~\ref{Fig:RSBoo}.}
\label{Fig:SUDra}
\end{figure}

\oc~variations of the star SU Dra cannot be explained by a simple parabolic trend. As in other mentioned cases, we proposed that period changes are caused by the presence of an unseen body. According to our model, combining parabola and \lt~(Fig.~\ref{Fig:SUDra}), the low-mass companion ($\mathfrak{M}_{\rm 2, min} \sim 0.09$\,$\mathfrak{M}_{\odot}$) should orbit SU Dra with a period of 70.7 years. The parabolic shape of \oc~suggests a period change rate of 16.6$^{+0.7}_{-1.2}\times 10^{-11}$\,d\,d$^{-1}$ which is almost the same value as 17.0(1.0)$\times 10^{-11}$\,d\,d$^{-1}$ \citep{leborgne2007}\footnote{He calculated also third-order fit for SU Dra with 13.4$\times 10^{-11}$\,d\,d$^{-1}$.}. An additional scenario (simple \lt) was tested and similar to other proposed systems, the found orbital period of about 149\,yr is very uncertain since only an incomplete cycle is available in the \oc~diagram.

SU Dra, as the brightest star from our sample, has the largest number of sources with available RV measurements (Table~\ref{Tab:RVPublicationtable}). Unfortunately several values are dubious and the remaining data seem to contradict a binary hypothesis. However, the semi-amplitude of predicted RV changes $K_{1}$ is quite small -- only 0.85\,km\,s$^{-1}$ which is hardly detectable. Therefore this test should not be considered as reliable.

% ===========================================================
% ===========================================================
\subsection{VX Her}\label{resultsvxher}

\begin{figure}
\centering
\includegraphics[width=0.95\hsize]{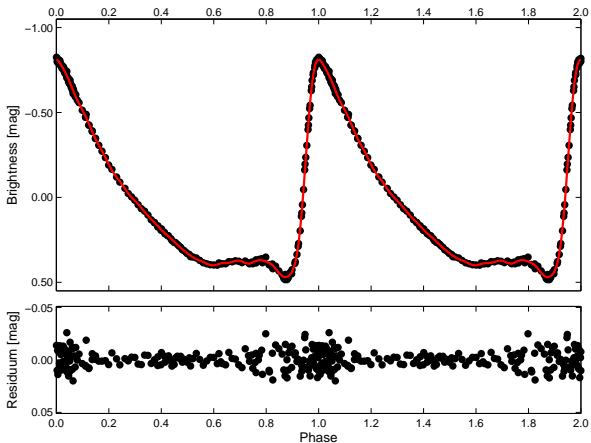}
\caption{Light curve of VX Her in $V$ band obtained in order to detect eclipses at MUO during 13 nights (April\,--\,August 2014) together with the polynomial model (top). The residual after model subtraction (bottom) shows no evident variations larger than 0.02\,mag.}
\label{Fig:VXHerLC}
\end{figure}

VX Her was proposed to be an eclipsing binary by \citet{fitch1966} -- see the discussion in Sect.~\ref{rreclsubsec} for more details. Unfortunately, no other observation from literature nor our observations (Fig.~\ref{Fig:VXHerLC}) confirm Fitch's results. 

\begin{figure}
\centering
\includegraphics[width=0.95\hsize]{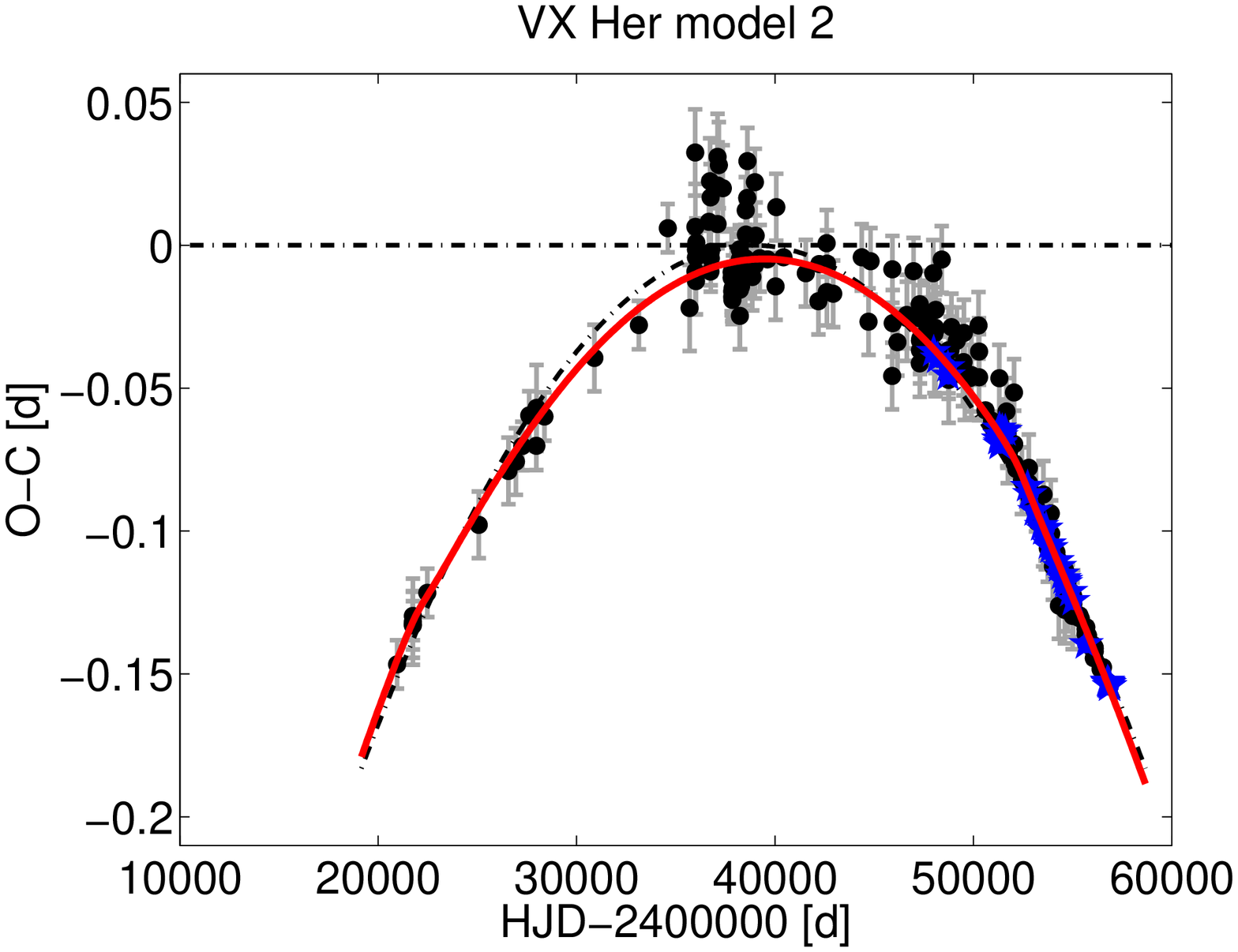}
\includegraphics[width=0.95\hsize]{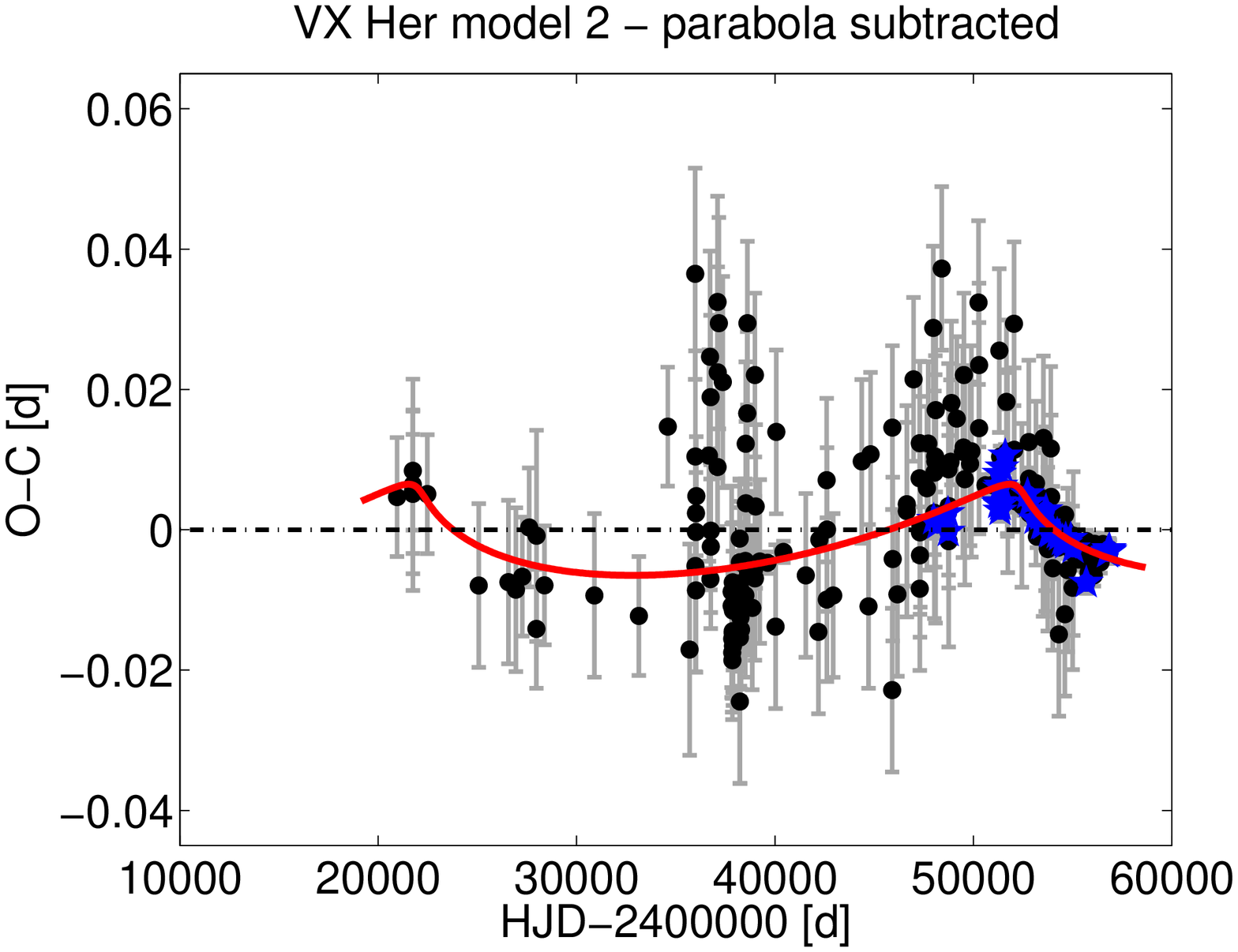}
\caption{\oc~diagram of VX Her together with our model 2 (top panel) and variation after subtracting the parabolic trend (bottom panel). Symbols are the same as in Fig.~\ref{Fig:RSBoo}.}
\label{Fig:VXHer}
\end{figure}

The shape of the \oc~diagram (the top panel of the Fig.~\ref{Fig:VXHer}) of VX Her is apparently a parabola with a small systematic deformation. The residuum after subtraction of the parabolic trend (bottom panel of the Fig. \ref{Fig:VXHer}) contains a small wave which can be described by a model of \lt~with a period of about 86\,yr. Nevertheless, the existence of \lt~is a bit questionable due to the high scatter in older maxima (photographic, visual). At least our result can be considered as a limiting value for \lt~in VX Her (the limiting mass of a second component could only be 0.05\,\M). The value found for the period change rate $-42.71^{+66}_{-48}\times10^{-11}$\,d\,d$^{-1}$ is similar to the value $-40.5(9)\times10^{-11}$\,d\,d$^{-1}$ from \citet{leborgne2007}. 

All the \oc~variations can also be described using only simple \lt~with a longer period. In such a highly unlikely scenario the possible orbital period would be about 175\,yr, and the limiting mass of the secondary star would then be about 0.8\,$\mathfrak{M}_{\odot}$.

According to \citet{azarnova1963} and \citet{wunder1990} the pulsations are modulated with a \Blazhko~period of 455.37\,d with quite a high amplitude of \oc~changes of 0.013\,d. This periodicity was found after subtraction of the parabola in the \oc~diagram without verification of light variations. Later results from Wunder are based only on photographic, visual and 4 photoelectric measurements of maxima timings and they should be verified. Moreover these changes may not be caused by the \Blazhko~effect, but by \lt. In our residuals after subtracting the parabola, which are based on photoelectric and CCD measurements, no indications of modulation are apparent. Since these residuals have a standard deviation of 0.0015\,d, changes in amplitude of a magnitude larger should definitely be detectable. Frequency analysis of residuals from photographic and visual measurements \citep[including measurements of][]{wunder1990} show no indications of periodicity. In addition, \Blazhko~modulation was undetectable in ASAS data \citep{skarka2014} and in our measurements (Fig.~\ref{Fig:VXHerLC}). Therefore it is very likely that VX Her is currently unmodulated.

RVs from two available sources (see Table~\ref{Tab:RVPublicationtable}) have insufficient quality to confirm the binarity, because the range of RV variations is from $-410$ to $-340$\,km\,s$^{-1}$, while our model suggesting a 0.05-\M~companion predicts the semi-amplitude of RV to be only 0.8\,km\,s$^{-1}$.

% ===========================================================
% ===========================================================
\subsection{SS Leo}\label{resultsssleo}

\begin{figure}
\centering
\includegraphics[width=0.95\hsize]{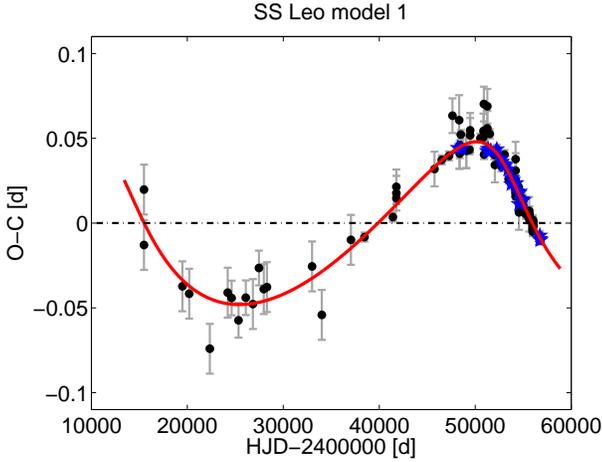}
\caption{\oc~diagram of SS Leo together with our model~1. Symbols are the same as in Fig.~\ref{Fig:RSBoo}.}
\label{Fig:SSLeo}
\end{figure}

The \oc~diagram of SS Leo shows large changes in the pulsation period (Fig.~\ref{Fig:SSLeo}) that can be explained by long-period \lt. Apparently it could not be explained as secular period change with additional low-amplitude \lt. We found a probable orbital period with length of about 111 years. Therefore, only one cycle was completed since the time of the first observation. SS Leo is known as a regular pulsating star without the \Blazhko~effect \citep{skarka2013}, and we did not either notice signs of the \Blazhko~effect. The proposed companion should have a minimum mass of 0.38\,\M.

Measurements of RV were found in four publications (see Table~\ref{Tab:RVPublicationtable}). Unfortunately, three of them contain values obtained over 4 years, thus practically in the same phase of the supposed orbit. During this time the difference in systemic velocity based on our \lt~model should change only within 0.16 km\,s$^{-1}$, and this difference is below the accuracy of these RV measurements. Nevertheless, the pulsation phase is well covered by RVs and values vary in the range of 120\,--\,205\,km\,s$^{-1}$. One of the values published by \citet{abt1973} ($-20$\,km\,s$^{-1}$) is very different from the others and is probably incorrect. Again it is impossible to confirm the binarity of SS Leo using available RV measurements.

However, in May 2015 the systemic velocity should be lower by about 4\,km\,s$^{-1}$ than for RV measurements from the last three sources. We propose SS~Leo as an easy target for confirmation of the binary nature by RV measurements.

% ===========================================================
% ===========================================================
\subsection{AV Peg}\label{resultsavpeg}

\begin{figure}
\centering
\includegraphics[width=0.95\hsize]{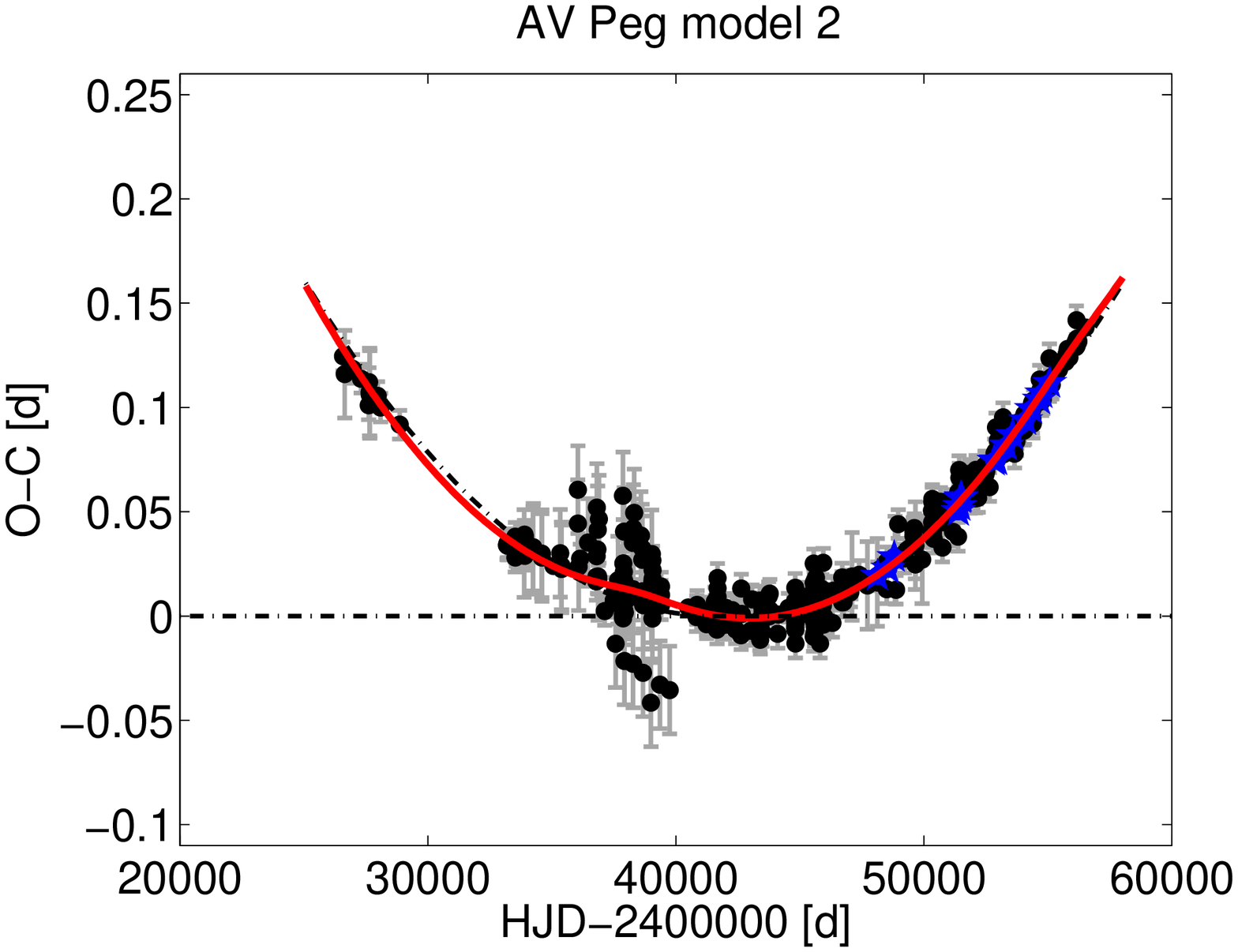}
\includegraphics[width=0.95\hsize]{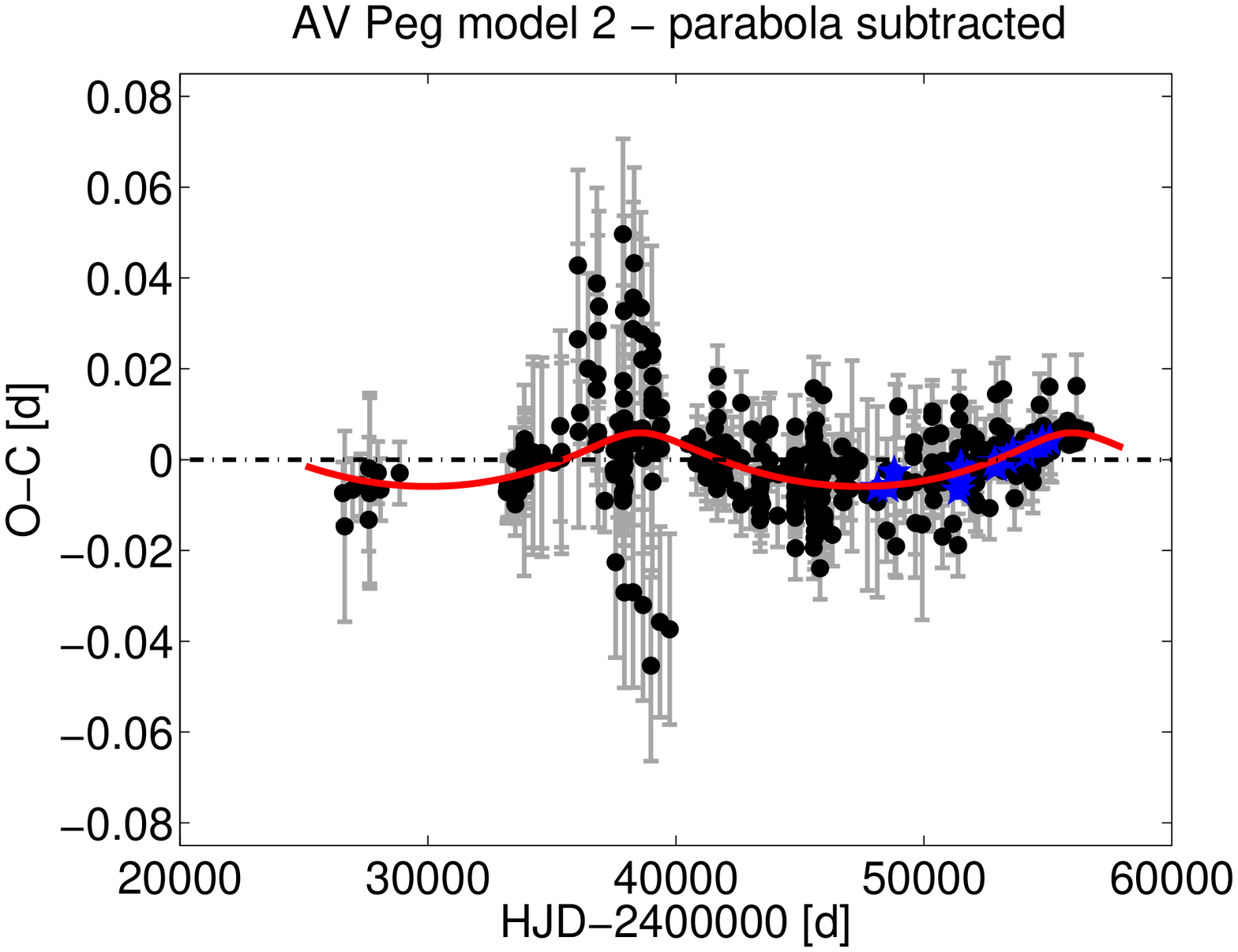}
\includegraphics[width=0.95\hsize]{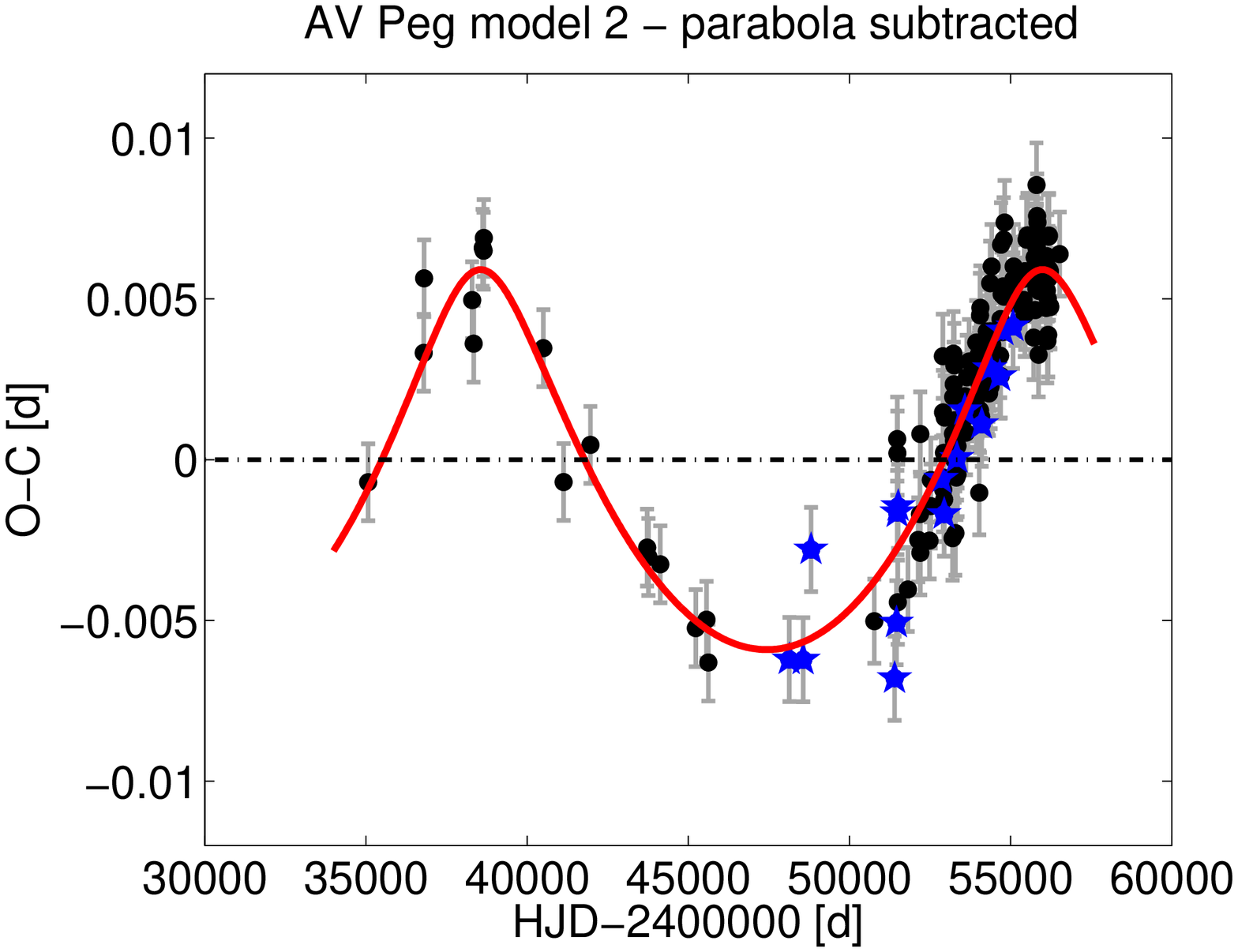}
\caption{\oc~diagram of AV Peg together with our model 2 (top panel). The middle panel demonstrating \lt~shows all data after subtracting the parabola, the bottom panel shows the same only for CCD and photoelectric measurements. Symbols are the same as in Fig.~\ref{Fig:RSBoo}.}
\label{Fig:AVPeg}
\end{figure}

AV Peg's \oc~diagram has a parabolic shape (Fig.~\ref{Fig:AVPeg}). In the residual, after applying a quadratic ephemeris, a wave was detected, which was proposed by \citet{szeidl1986} as a sign of \lt. Their preliminary value of the orbital period was estimated to be 64 years. Our approach, in which parameters of a quadratic ephemeris and \lt~ are fitted simultaneously (top panel of Fig.~\ref{Fig:AVPeg}), yields a shorter orbital period of 48\,yr and the rate of secular change of the pulsation period $\dot{P}_{\rm puls}=+45.96^{+22}_{-42}\times 10^{-11}$\,d\,d$^{-1}$ which is very similar to values $+47.3(8)\times 10^{-11}$\,d\,d$^{-1}$ \citep{szeidl1986} and  $+45.8(5)\times 10^{-11}$\,d\,d$^{-1}$ \citep{leborgne2007}. The difference in the length of the orbital period can be explained by using about a 30 year longer dataset with more accurate CCD measurements than was used by \citet{szeidl1986}. Our model of \lt~in AV Peg gives the second lowest minimum mass of the secondary component of all stars in our sample (only 0.06\,\M). When applying the model only to CCD and photoelectric measurements and subtracting the parabolic trend (bottom panel of the Fig.~\ref{Fig:AVPeg}), the \lt~is more pronounced than in the case of all measurements (middle panel of the Fig.~\ref{Fig:AVPeg}).

RVs found in 3 sources (Table~\ref{Tab:RVPublicationtable}) were obtained in phases where the systemic velocity should differ by only about 0.26\,km\,s$^{-1}$. Thus the variations in RV due to binary motion are undetectable in these measurements. The range of RVs due to pulsation of the star\footnote{One value from \citet{abt1973} substantially differs from the others.} is from $-93$ to $-31$\,km\,s$^{-1}$. In any case, the semi-amplitude of RVs caused by the proposed orbit is too low (only 0.7\,km\,s$^{-1}$) and will be very difficult to confirm.

% ===========================================================
% ===========================================================
\subsection{AT Ser}\label{resultsatser}

\begin{figure}
\centering
\includegraphics[width=0.95\hsize]{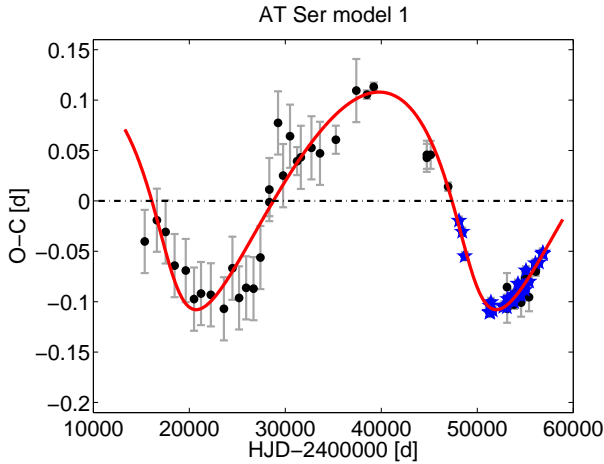}
\caption{The \oc~diagram of AT Ser together with our model~1. Symbols are the same as in Fig.~\ref{Fig:RSBoo}.}
\label{Fig:ATSer}
\end{figure}

More than one orbital cycle is covered for times of maximum for AT~Ser (Fig. \ref{Fig:ATSer}). From the amplitude of changes in \oc~it is evident that the possible companion has a high mass. Our model, with an orbital period of 86 years, predicts a minimum mass of the second component of 1.9\,\M. Therefore the companion should be a degenerate remnant in the form of a neutron star, or more likely in the form of a black hole.  

RVs were found in 3 sources (Table~\ref{Tab:RVPublicationtable}). Unfortunately, these 9 values do not even cover a complete pulsation phase. Measured RVs vary between $-100$ and $-40$\,km\,s$^{-1}$, but the full range of variations in velocity will be a bit larger. The calculated semi-amplitude of the orbital RV curve for this system is one of the largest -- about 7.8\,km\,s$^{-1}$. Currently (May 2015), the systematic offset in RVs should have the highest difference of about 15.5 km\,s$^{-1}$ other than for RVs from \citet{fernley1993}. AT Ser is, therefore, another easy target for proving binarity using RVs.

% ===========================================================
% ===========================================================
\subsection{RV UMa}\label{resultsrvuma}
\begin{figure}
\centering
\includegraphics[width=0.95\hsize]{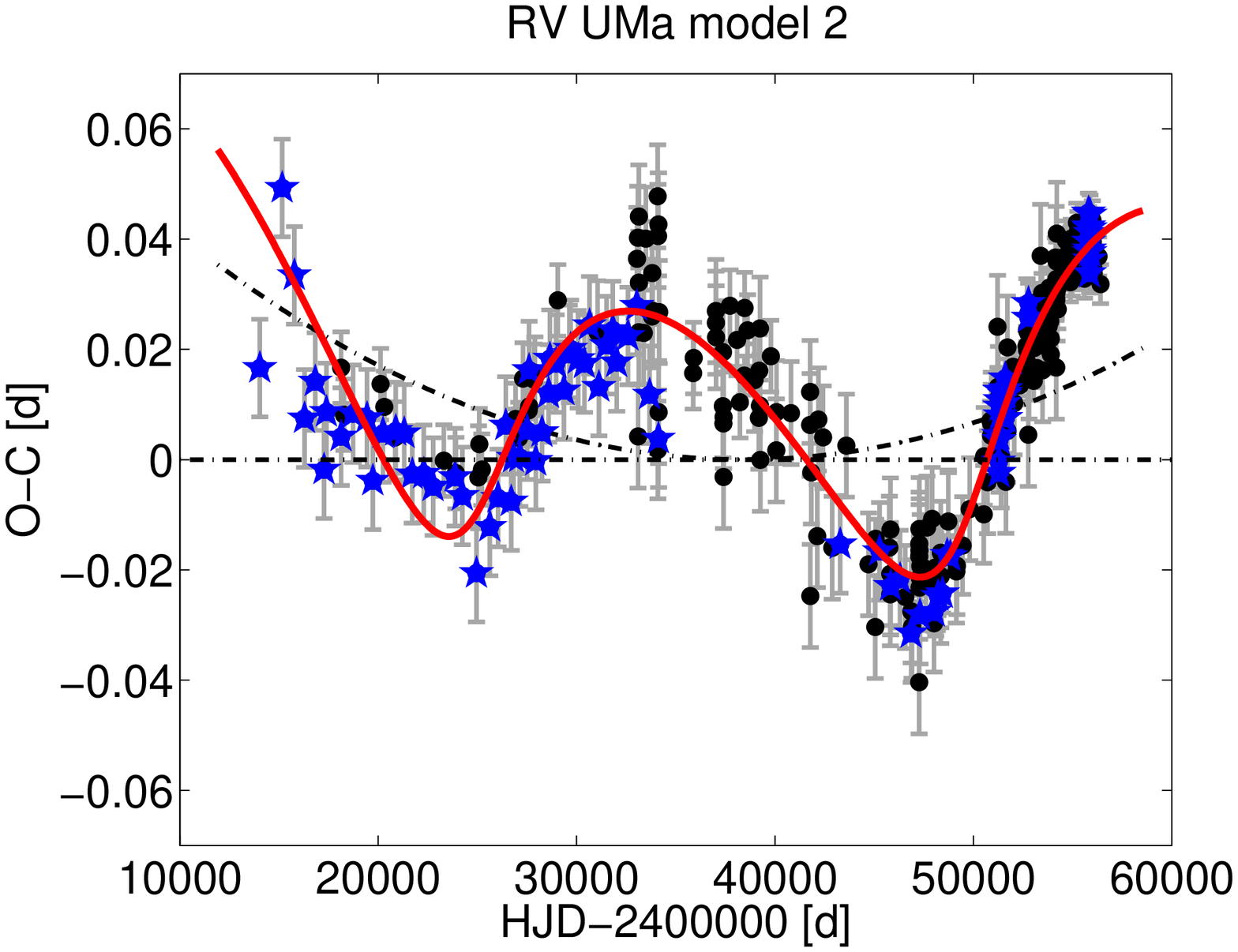}
\includegraphics[width=0.95\hsize]{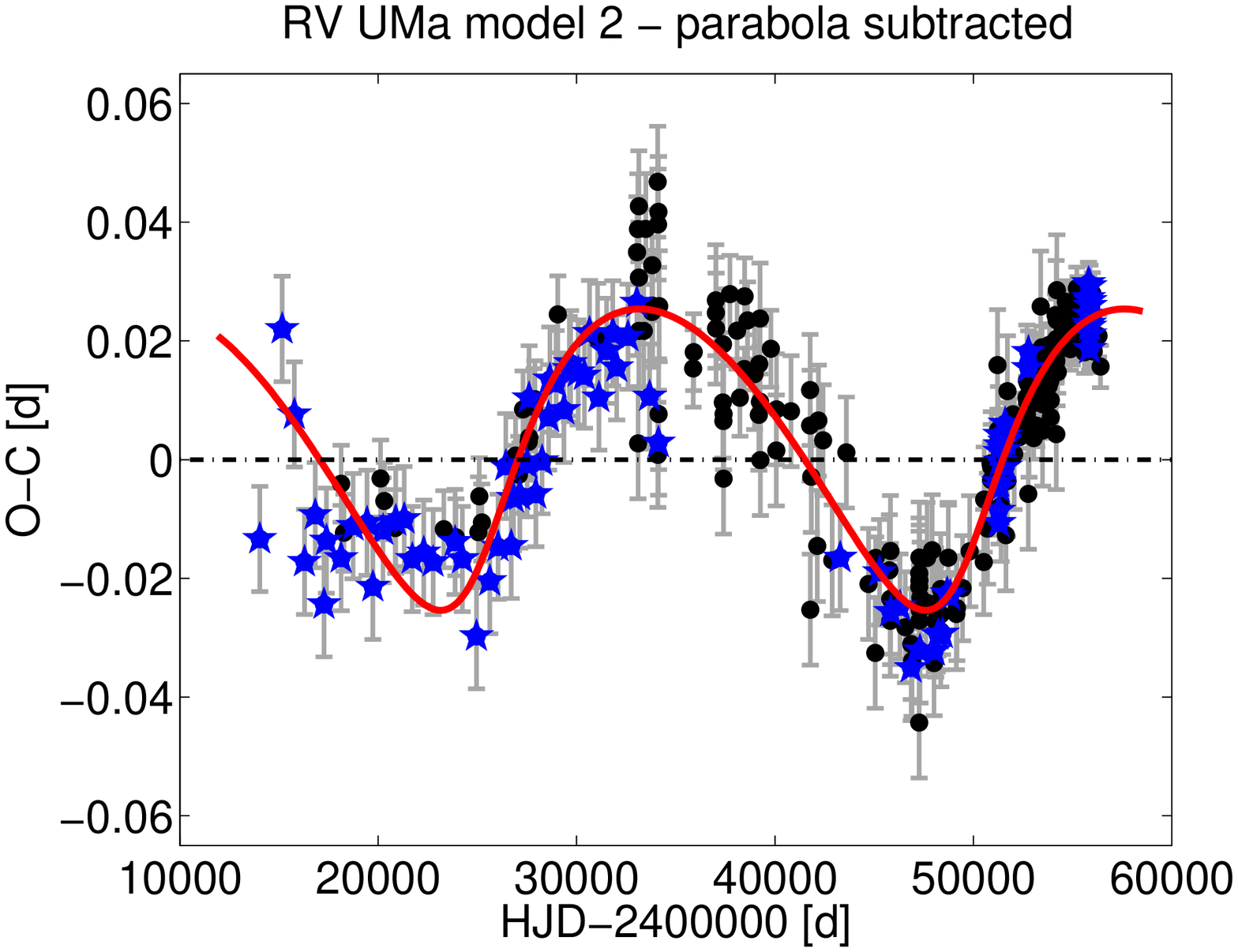}
\caption{\oc~diagram of RV UMa together with our model 2 (top panel) and variation after subtracting the parabolic trend (bottom panel). Symbols are the same as in Fig.~\ref{Fig:RSBoo}.}
\label{Fig:RVUMa}
\end{figure}

RV UMa is a well known modulated star with secular period changes accompanied with \lt~covering more than one proposed orbital cycle (Fig. \ref{Fig:RVUMa}). When the \oc~dependence is considered as \lt~with a period of 66.9\,yr supplemented by a slow evolutionary period change (parabolic trend with the rate of $\dot{P}_{\rm puls}=4.7\times10^{-11}$\,d\,d$^{-1}$, the companion should have a minimum mass of 0.25\,$\mathfrak{M}_{\odot}$.

\begin{figure}
\centering
\includegraphics[width=0.95\hsize]{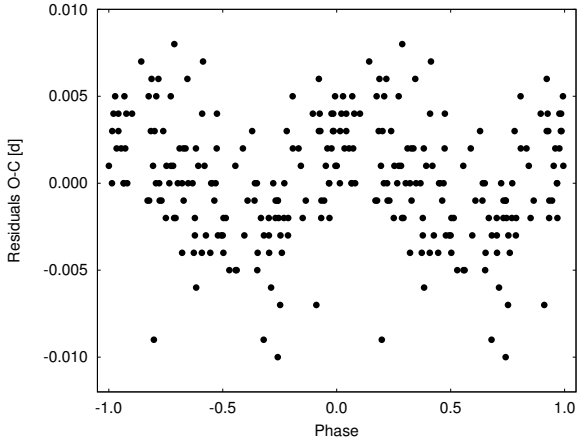}
\caption{\oc~residuals of RV UMa phased with the modulation period of 89.9(1)\,d.}
\label{Fig:RVUMaPhased}
\end{figure}

Except for these large-amplitude period changes, RV UMa shows additional small changes caused by the \Blazhko~effect with interesting behaviour. \citet{kanyo1976} analysed the period behaviour of RV UMa and found that the \Blazhko~period (about 90\,d) changes in the antiphase with the pulsation period. A more detailed analysis of period evolution and the \Blazhko~effect, performed by \citet{hurta2007} and \citet{hurta2008}, confirmed Kanyo's results and complemented it with finding that between 1946 and 1975 the changes in the pulsation and modulation periods were parallel. In addition, \citet{hurta2008} gives the range of the modulation-period changes between 89.9 and 90.6\,d\,\footnote{\citet{kanyo1976} found that in 1961 the \Blazhko~period was 92.12\,d.}. Nevertheless, none of these authors discussed the possible binary nature of the cyclic variations of the \oc~diagram shown in the Fig.~\ref{Fig:RVUMa}. After removing \lt~and the parabolic trend we searched for periodicity caused by the \Blazhko~effect. We found a modulation period of 89.9(1)\,d (Fig.~\ref{Fig:RVUMaPhased}). The high scatter probably corresponds to the variation in the length of the \Blazhko~period.

We attempted to confirm variations in the \oc~diagram by RV variations. Three datasets of RV measurements that cover the pulsation cycle well were found in literature (see Table~\ref{Tab:RVPublicationtable}). The measured values of RV were compared with RV values corrected for systemic velocity caused by orbital motion (parameters from our model of \lt). Unfortunately, confirmation failed due to a small semi-amplitude of $K_{1}$ of 2.2\,km\,s$^{-1}$ and due to changes in RV during the \Blazhko~effect -- \citet{preston1967} clearly presented evolution of the RV curve for RV UMa in different \Blazhko~phases (change of the shape and amplitude of more than 20\,km\,s$^{-1}$).

Our results, together with those from \citet{kanyo1976}, \citet{hurta2007} and \citet{jurcsik2002}, suggest that there could be some connection between orbital motion and variations in pulsation and \Blazhko~periods. Although \citet{hurta2007} did not deal with binarity, 67-yr antiparallel pulsation and modulation period changes, apparent from his Fig.~4, indicate this possibility. Similar but parallel behaviour was observed in XZ~Dra \citep{jurcsik2002}, and we assume that the connection with binarity should also be investigated in other \Blazhko~stars, e.g. XZ~Cyg \citep{lacluyze2004}, RW~Dra \citep{firmanyuk1978}. 

% ===========================================================
% ===========================================================
\subsection{BB Vir}\label{resultsbbvir}

\begin{figure}
\centering
\includegraphics[width=0.95\hsize]{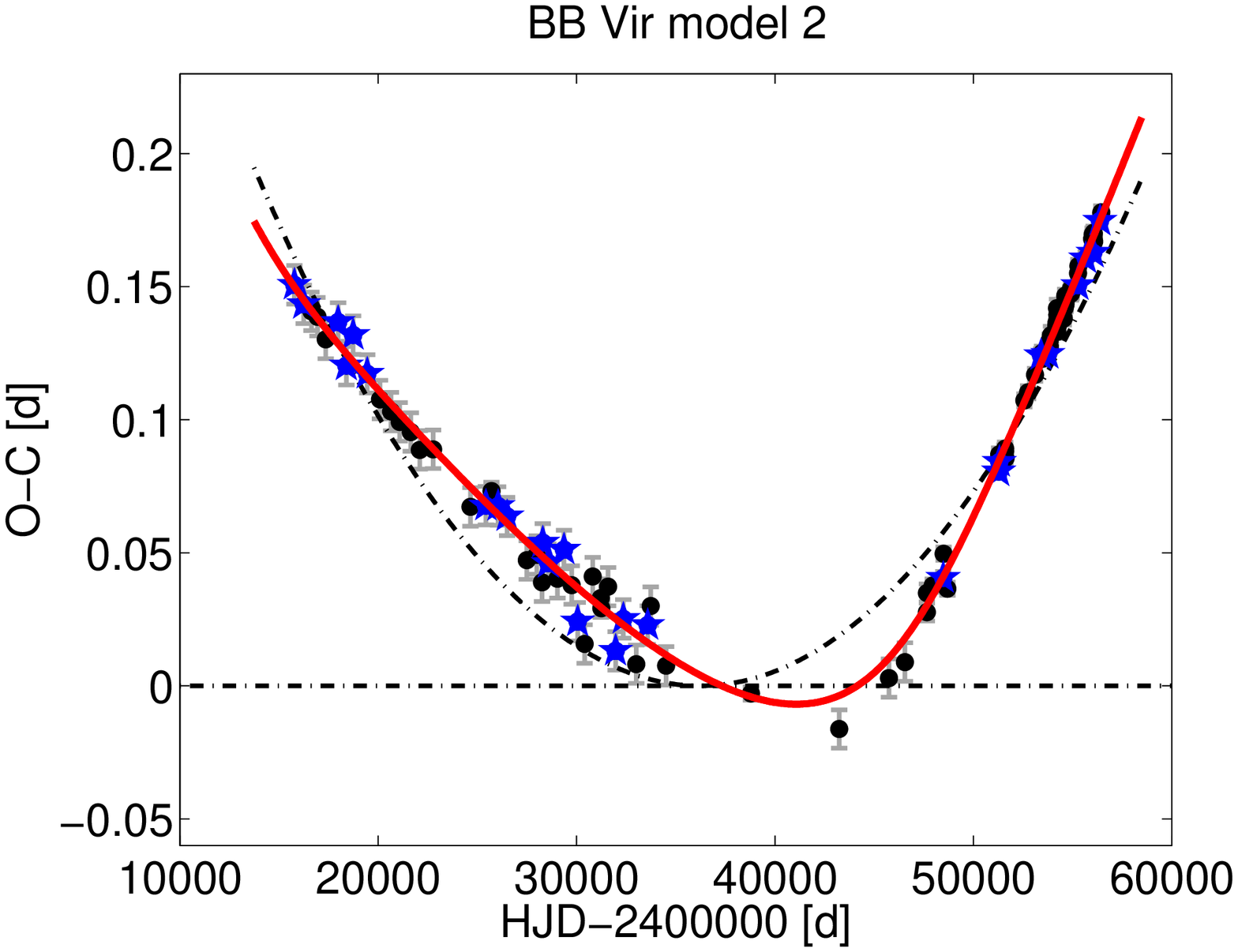}
\includegraphics[width=0.95\hsize]{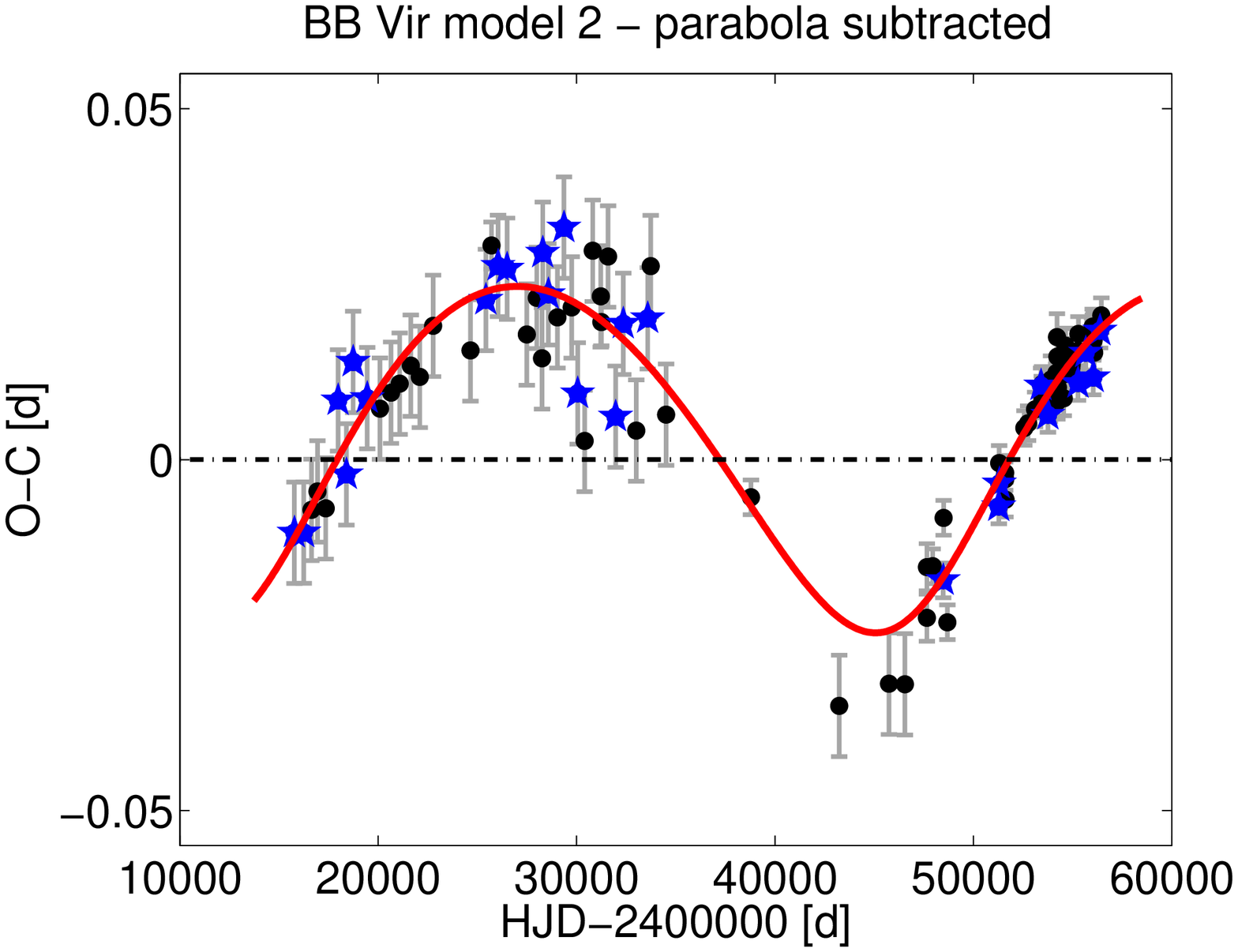}
\caption{\oc~diagram of BB Vir together with our model 2 (top panel) and variation after subtracting the parabolic trend (bottom panel). Symbols are the same as in Fig.~\ref{Fig:RSBoo}.}
\label{Fig:BBVir}
\end{figure}

\citet{kinman1992} supposed BB Vir to be a member of a binary with a hot HB star (already discussed in Sect.~\ref{rrothersec}). When we check Fig.~\ref{Fig:BBVir} we immediately see that the \oc~dependence is deformed and could not be fitted with a simple parabola. One of the possible explanations is that the \oc~is a part of the \lt~with an incomplete cycle. Assuming this less probable scenario, we found that the companion with a minimum mass of about 0.95\,\M~should orbit around the RR Lyrae component at a minimum distance of 21.6\,au with an orbital period of about 168 years. The relatively high minimum mass suggests a degenerate remnant.

Nevertheless, the second model combining the parabolic trend with \lt~(Fig. \ref{Fig:BBVir}) is supposed to be more probable. We found a period of about 93\,yr and the rate of period changes $\dot{P}_{\rm puls}=+36.3^{+1.5}_{-0.3}\times 10^{-11}$\,d\,d$^{-1}$. The obtained period-change rate differs from value $+46.0(3.8)\times 10^{-11}$\,d\,d$^{-1}$ \citep{leborgne2007} which can be explained by adding \lt~to our model.  

Because BB Vir is known as a star with modulation in the GCVS catalogue, we unsuccessfully searched for additional periodicity in residuals. In addition, the modulation period is not known, and no signs of modulation were found in ASAS data by \citet{skarka2014}. Therefore the \Blazhko~effect in BB Vir is at least questionable.

Only one single RV measurement $-15.6$\,km\,s$^{-1}$ \citep{abt1973} is available in literature, but it is insufficient to tell us anything about the binarity of BB Vir.

% ===========================================================
% ===========================================================
\subsection{Other RR Lyrae stars with long-term cyclic variations in \oc~mimicking \lt}\label{resultsotherstars}

\begin{figure}
\centering
\includegraphics[width=0.95\hsize]{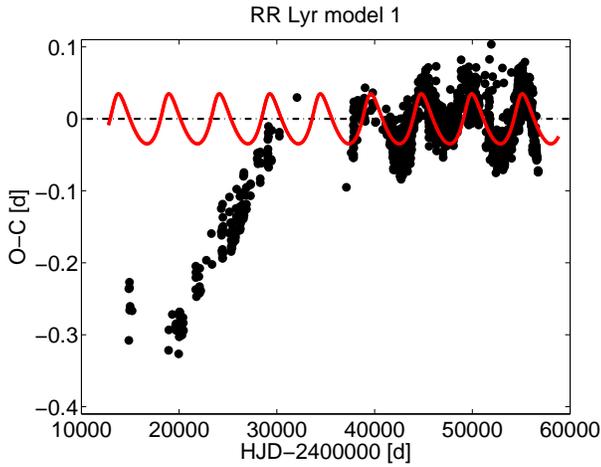}
\caption{\oc~diagram of RR Lyrae (after the gap at about JD 2435000) shows also cyclic changes which could be described by the \lt~model.}
\label{Fig:RRLyr}
\end{figure}

\begin{figure}
\centering
\includegraphics[width=0.95\hsize]{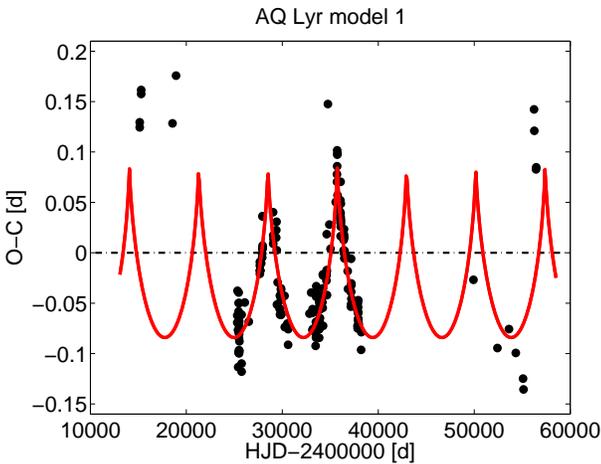}
\caption{\oc~diagram of AQ Lyrae that shows cyclic changes during a short interval (JD 2424000\,--\,2438000) which could be described by the \lt~model.}
\label{Fig:AQLyr}
\end{figure}

There could be several other RR Lyrae type stars that exhibit cyclic changes that could possibly be interpreted through \lt. Nevertheless, the shape of variations often changes from cycle to cycle which is difficult to explain by simple \lt. A typical example is the star RR~Lyr itself. Its semi-regular variations after the gap at about JD 2435000 resemble the \lt~with a 14.1-yr period (Fig.~\ref{Fig:RRLyr}) and a companion with a minimum mass of 2\,\M~with an eccentric orbit with $e\sim0.4$. Unfortunately, the general trend in period evolution of RR Lyrae is more complex with the gap, and different slopes before and after the gap. The situation is similar with AQ Lyr (Fig.~\ref{Fig:AQLyr}) which underwent two cycles during the interval JD 2424000\,--\,2438000 with a period of 19.8\,yr well approximated by a highly eccentric orbit ($e\sim0.75-0.95$) of a massive component 6\,--\,9\,\M. Before and after this interval the star behaved differently.

\begin{figure}
\centering
\includegraphics[width=0.95\hsize]{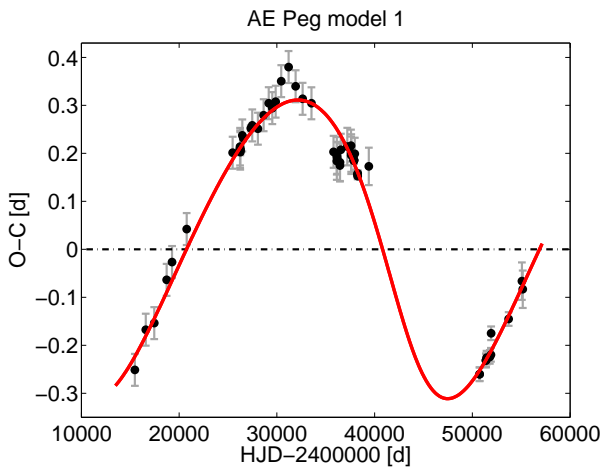}
\caption{\oc~diagram of AE Peg that shows probable cyclic changes which could be described by the \lt~model.}
\label{Fig:AEPeg}
\end{figure}

Another candidate for binarity is AE Peg showing high amplitude variations in \oc~(Fig.~\ref{Fig:AEPeg}) with a period of about 99 years, $e\sim 0.3$, and a semi-major axis $a_{1}\,\sin i \sim 56$\,au. The secondary component should be a massive black hole with a mass of about 18.6\,\M. Maybe a binary explanation is not correct for these objects, but hypotheses employing other effects discussed in Sect.~\ref{rrocsec} are equally uncertain.

These three mentioned objects have known \Blazhko~modulations with lengths of tens of days. Detected long-term cyclic variations can be a manifestation of multiple \Blazhko~effects. These objects are good examples of stars with complex behaviour that looks like \lt~in some parts, but in others are undetectable. In the case of AQ~Lyr and RR~Lyr, a time-base longer than 2 and 4 cycles, respectively, was necessary to detect a completely different kind of variations in the \oc~diagram. Additional observations will be highly appreciated.

%*************************************************************************
\section{Discussion}\label{discussionsec}
%*************************************************************************
Cyclic changes in \oc~diagrams can be interpreted in various ways. The time scale of the changes that we studied is crucial. Thus far the longest observed process in RR~Lyrae stars causing cyclic changes (not necessarily strictly periodic) in the \oc~diagram, is the \Blazhko~effect. Observed periods are maximally in the order of years, while we analysed period-evolution in the order of several decades. Another possibility to explain cyclic long-term period variations could be the presence of a solar-like magnetic cycle. This idea was proposed by \citet{stothers1980}. \citet{derekas2004} argued that the relative change in radius during the period-change cycle of BE~Dor is in agreement with the relative change of solar radius during its magnetic cycle \citep{noel2004}, and therefore a hydromagnetic interpretation could be correct. The problem is that a dipole magnetic field in RR~Lyrae stars was not detected \citep[e.g.,][]{chadid2004,kolenberg2009}, and it is not clear whether solar-like magnetism is even possible in pulsating HB stars. Perhaps some unknown physical mechanism could be responsible for the long-term cyclic evolution of the pulsation period. However, the most simple, and currently most probable explanation is through binarity, which we assume in this paper.  

Our target selection gives preference to orbits with wide trajectories and with long orbital periods lasting for many decades. Such systems are, therefore, definitely well detached without mass transfer between components taking place during their life cycles. This should allow direct determination of the true RR Lyrae mass which is not influenced by stellar evolution in a close binary system -- a chance keenly awaited by all theoreticians dealing with stellar evolution and pulsations. However, wide trajectories make eclipses highly improbable, not even mentioning eclipse prediction, which strongly degrades the chance of precisely determining the mass of an RR Lyrae component in this way. Observation of eclipses is therefore a very unlikely event. Maybe the dimming of the light of VX Her noted by \citet{fitch1966} was such a case. In spite of the slim chance for eclipse detection, wide trajectories can be used with the advantage of determining RR Lyrae mass on the basis of visual binary motion together with RV measurements.

However, the recent studies by \citet{li2014} and \citet{hajdu2015} detected systems with much shorter orbits only in the order of years\footnote{Binary system TU UMa \citep{liska2015} with a 23.4\,yr period is in the middle of the orbital period interval of the mentioned survey systems and those from this paper.} giving a better chance for catching an eclipse. OGLE and {\it Kepler} measurements span maximally a few years. Thus long-period binaries remained hidden in these observations. A great advantage of the data from OGLE and space surveys against our sample is that they observed many thousands of targets almost continuously, while we analysed bright field targets with a large amplitude of variations in their \oc, that are sparsely observed separately by different observers, and not regularly and with similar cadence as in the case of Kepler or OGLE. The time-span of the data that \citet{li2014} and \citet{hajdu2015} used, caused that their results are strongly biased towards short orbital periods.

Nevertheless, a project established for period studying of RR Lyraes, using their maxima timings regularly and densely observed by the TAROT telescopes \citep{leborgne2004,leborgne2007}, can be considered as comparable to the OGLE. Detailed analysis of all \oc~diagrams (not only a visual selection as was done in this paper) for these TAROT objects would reveal systems with periods of several years.

Another problem which relates to the length of the proposed orbital periods, are the uncertainties surrounding the choice for modelling of the \oc s (with or without secular parabolic trends). \citet{hajdu2015} applied secular changes together with \lt~by default. Generally, our models with parabolic trend give much shorter periods than models with \lt~alone. 

Companions of all stars in our sample should be objects with minimum mass larger than about 0.05\,\M. In the case of VX Her and AV Peg, their companions could be brown dwarfs when the orbits would have an inclination close to 90 degrees. In other examples the companions are probably low-luminous low-mass main sequence red dwarfs, or stellar remnants. This corresponds well with the absence of information about the peculiar colour of any of the stars in literature\footnote{\citet{kinman1992} found BB Vir to be somewhat bluer than other RR Lyrae stars.}. If the companion was a luminous star of a different colour than the RR Lyrae component, we would observe significant colour excess in comparison with the average colour of RR~Lyraes.

In RZ Cet and AT Ser our models suggest high-mass companions in the form of a degenerate remnant with a minimum mass of about 1.15\,\M~and 1.90\,\M, respectively. Progenitors of these massive remnants produced heavier elements without a doubt and definitely should enrich their RR Lyrae partners with metals -- as it was proposed by \citet{kennedy2014}. Thus observation of higher metal abundances should be an independent and relatively easy way to confirm the binary nature of RZ Cet and AT Ser.

In addition, the wide orbits of our systems represent a perfect opportunity to prove their binarity through direct imaging. The distance, calculated using the observed and average absolute magnitude (0.6\,mag)\footnote{Hipparcos parallaxes are unreliable for the studied stars.} together with a computed semi-major axis, gives an angular distance between binary components ranging from 1 to 13\,mas for our sample stars. This is about ten times more than for stars in the Galactic bulge studied by \citet{hajdu2015}. In addition, the appropriate orientation of the orbit can further double these numbers. On the other hand, the angular distance could be much lower. The resolution of a few mas should be reachable by state-of-the-art ground-based and space instruments. Because the proposed companions are low-mass, low-luminous red stars, the detection would be less demanding in the red part of the spectrum where the difference between members of the system is less. However, even in IR the difference would be several magnitudes which could cause some problems in detection. The most promising candidates with the largest semi-major axes are AT~Ser (13\,mas), RS Boo (9\,mas, model 1) and RZ~Cet (8\,mas).

We did not notice any correlation between the presence of the \Blazhko~effect and binarity. However, some unknown mechanism connected with binarity can induce variation in the length of the \Blazhko~period as was discussed for RV UMa in Sect.~\ref{resultsrvuma}. Also metallicity probably does not influence the fraction of RR Lyraes in binary stars.

\section{Summary}\label{summarysec}

In this study we focused on RR Lyrae type stars with suspicious \oc~diagrams showing cyclic variations or some indications of possible long-term periodic changes, preferably of large amplitude. We attribute them to the consequence of binarity manifesting itself as the Light Time (Travel) Effect. We attempt to provide an extensive overview of known, or more accurately, suspected RR Lyrae type pulsators in binaries, and we discuss problems involving observational and analytical methods for the confirmation of the binarity. We also widely discuss various effects which can influence the appearance of \oc~diagrams to establish a solid basis for interpretation and strengthening of our results.

We proceeded from the collection of maxima timings provided in the GEOS RR Lyrae database \citep{leborgne2007} and selected several stars with \oc~diagrams whose appearance could be interpreted as the consequence of binarity. In addition to data from the GEOS database, we determined maxima timings for selected targets from large sky surveys and our observations. For the modelling of the \lt, we used recently developed code based on the non-linear LSM and bootstrap-resampling method \citep{liska2015}. \lt~for eleven stars were modelled and possible orbital parameters were calculated (Table~\ref{Tab:LiTEtable}). Orbital periods of hypothetical binaries range from 47 (AV Peg) to 147 years in RS Boo (model 1). The projection of the semi-major axis of the pulsating component $a_{1}\sin i$ was estimated as 1\,au in AV Peg and about 20\,au in AT~Ser. Minimum masses of companions were obtained according to our expectation (see Sect.~\ref{introductionsec}). They were either very small (hundredths to tenths of $\mathfrak{M}_{\odot}$) or higher than the solar mass. All proposed systems have eccentric orbits, but none of them has $e>0.9$. The majority of studied stars has an eccentricity between 0.1 and 0.5. Only RS Boo and VX~Her could have more eccentric orbits.

Although in many targets we could not clearly decide which model of \oc~variations is correct (only \lt, or \lt~superimposed on a parabola), our models provide a prediction for the period evolution in future years, as well as for RV which could be expected to be observed in years to come. Consequently only the future can provide a reliable answer to the question whether RR Lyrae stars analysed in this study are really bound in binary systems or not. Nevertheless, our study substantially extends the group of suspected RR Lyrae variables in binary systems with preliminary, estimated orbital parameters.

%*************************************************************************
\section*{Acknowledgements}
%*************************************************************************
Based on data from the OMC Archive at CAB (INTA-CSIC), pre-processed by ISDC. The DASCH project at Harvard is grateful for partial support from NSF grants AST-0407380, AST-0909073, and AST-1313370. This paper makes use of data from the DR1 of the WASP data \citep{butters2010} as provided by the WASP consortium, and the computing and storage facilities at the CERIT Scientific Cloud, reg. no. CZ.1.05/3.2.00/08.0144 which is operated by Masaryk University, Czech Republic. We thank people from GEOS for maintaining their RR Lyr database. This work was supported by the grants MUNI/A/1110/2014, MUNI/A/1494/2014 and LH14300. MS acknowledges the support of the postdoctoral fellowship programme of the Hungarian Academy of Sciences at the Konkoly Observatory as a host institution. Our thanks go to several persons for their help: the anonymous
referee, our colleague R.~F.~Auer, and people from libraries, namely to Martina Antlov\'{a} and Kate\v{r}ina Sold\'{a}nov\'{a}.

%__________________________________________________________________

%\begin{appendix}\label{appendix1}
%\newpage
%\onecolumn
%\section{New maxima timings for pulsating star XX XXX}\label{maximaappendix1}
%\begin{center}
%\begin{table}%[htbp]
%\centering
%{\tiny
%\def\arraystretch{1.5}
%\tabcolsep=1.3pt
%\begin{tabular}{cccc|ccccc}
%\hline\hline
%$T_{\rm max}$ [HJD]	& error [d]	& Project & Method & $T_{\rm max}$ [HJD]	& error [d]	& Project & Method & Notes\\
%\hline
%2414783.4832 & 0.0020 & DASCH & pg & 2432312.9088 & 0.0037 & DASCH & pg &\\                      
%\hline
%\end{tabular}}
%\end{table}
%\end{center}

%\end{appendix}

\end{document}